\documentclass{aastex631}

\usepackage{color}
\usepackage{bm}
\usepackage{amsmath}

\begin{document}

\title{Radiation MHD Simulations of Soft X-ray Emitting Regions in Changing Look AGN}

\author{Taichi Igarashi}
\affiliation{Division of Science, National Astronomical Observatory of Japan, 2-21-1 Osawa, Mitaka, Tokyo 181-8588, Japan}
\affiliation{Department of Physics, Rikkyo University, 3-34-1 Nishiikebukuro, Toshima-Ku, Tokyo 171-8501, Japan}

\author{Hiroyuki R. Takahashi}
\affiliation{Department of Natural Science, Faculty of  Arts and Science, Komazawa University, 1-23-1 Komazawa, Setagaya-Ku, Tokyo 154-8525, Japan}

\author{Tomohisa Kawashima}
\affiliation{Institute for Cosmic Ray Research, The University of Tokyo, 5-1-5, Kashiwanoha, Kashiwa, Chiba 277-8582, Japan}

\author{Ken Ohsuga}
\affiliation{Center for Computational Sciences, University of Tsukuba, Tennodai, 1-1-1, Tsukuba, Ibaraki 305-8577, Japan}

\author{Yosuke Matsumoto}
\affiliation{Institute for Advanced Academic Research, Chiba University, \\
1-33 Yayoi-Cho, Inage-Ku, Chiba 263-8522, Japan}

\author{Ryoji Matsumoto}
\affiliation{Digital Transformation Enhancement Council, Chiba University, \\
1-33 Yayoi-Cho, Inage-Ku, Chiba 263-8522, Japan}



\begin{abstract}
Strong soft X-ray emission called soft X-ray excess is often observed in luminous active galactic nuclei (AGN).
It has been suggested that the soft X-rays are emitted from a warm ($T=10^6\sim10^7\ \rm{K}$) region that is optically thick for the Thomson scattering (warm Comptonization region). 
Motivated by the recent observations that soft X-ray excess appears in changing look AGN (CLAGN) during the state transition from a dim state without broad emission lines to a bright state with broad emission lines, we performed global three-dimensional radiation magnetohydrodynamic simulations assuming that the mass accretion rate increases and becomes around $10$\% of the Eddington accretion rate.
The simulation successfully reproduces a warm, Thomson-thick region outside the hot radiatively inefficient accretion flow near the black hole.
The warm region is formed by efficient radiative cooling due to inverse Compton scattering.
The calculated luminosity $0.01L_{{\rm Edd}}-0.08L_{{\rm Edd}}$ is consistent with the luminosity of CLAGN.
We also found that the warm Comptonization region is well described by the steady model of magnetized disks supported by azimuthal magnetic fields. 
{When the anti-parallel azimuthal magnetic fields supporting the radiatively cooled region reconnect around the equatorial plane of the disk, the temperature of the region becomes higher by releasing the magnetic energy transported to the region.}
\end{abstract}



\section{Introduction} \label{sec:intro}
Luminous active galactic nuclei (AGN), such as Seyfert galaxies and quasars, exhibit soft X-ray emission.
The origin of the soft X-ray emission in AGN is puzzling since the temperature of the standard accretion disk around a supermassive black hole is around $10^5\ \rm{K}$ \citep{shakura+sunyaev1973}.
Two types of models have been proposed to explain the soft X-ray emission.
One is the ionized reflection model \citep[e.g.,][]{gierlinski+done2004,garcia+2019}.
In this model, the soft X-ray emission is emitted by an optically thick disk that is irradiated by the hard X-ray emission from the vicinity of the central black hole. 
Another model is the emission from the Thomson thick warm corona with temperature $T=10^6-10^7\ \rm{K}$  by inverse Compton scattering \citep[e.g,.][]{magdziarz+1998,done+2012}.
\citet{kubota+done2018} showed that the intensity of the soft X-ray emission increases with increasing the mass accretion rate.
However, the geometry of the warm Comptonization region is still unknown, and it is not clear whether it is a region separated from the cool optically thick disk or associated with the cool disk.

Recent monitoring observations have shown that some Seyfert galaxies exhibit spectral state transitions 
\citep[e.g.,][]{shappee+2014,ricci+2020}.
They show transitions between a state without broad emission lines in the optical range (type 2 AGN) and a state with broad emission lines (type 1 AGN) {\citep[e.g.,][]{shapovalova+2019,oknyansky+2020,popovic+2023}.}
During the state transitions, the UV to X-ray spectrum also shows transitions {\citep[e.g.,][]{ricci+2021}.}
AGN that shows state transitions between type 1 and type 2 are called changing look AGN \citep[CLAGN; see][]{ricci+trakhtenbrot2023}.
\citet{noda+done2018} and \citet{tripathi+dewangan2022} showed that UV and soft X-ray emission is dominant in the bright phase of CLAGN, while hard X-ray emission is dominant in the dark phase.
During the state transitions, the soft X-ray intensity changes drastically.
Therefore, CLAGN may provide clues to the origin of the soft X-ray emission in AGN.

The spectral state transitions in CLAGN in the UV to X-ray range are similar to the hard-to-soft/soft-to-hard state transitions in black hole X-ray binaries (BHXB) observed during an outburst \citep[e.g.,][]{mcclintock+remillard2006,remillard+mcclintock2006,done+2007,belloni+motta2016}.
During the hard-to-soft state transitions, BHXB remains in a bright but hard X-ray dominant state over $\sim100$ days \citep[e.g.,][]{tanaka+shibazaki1996,shidatsu+2019}.
This state is called the bright-hard state.
In this state, the cut-off energy of the X-ray is anti-correlated with its luminosity.
It indicates that electrons 
temperature decreases as the accretion rate increases \citep{miyakawa+2008,motta+2009}.

To investigate the origin of the bright-hard state in BHXB, \citet{machida+2006} conducted three-dimensional global magnetohydrodynamic (MHD) simulations of BH accretion flows by including optically thin radiative cooling. 
{
The initial state is a hot magnetized radiatively inefficient accretion flow \citep[RIAF; ][see also \citealp{ichimaru1977}]{narayan+yi1995}.
}
When the disk surface density exceeds the upper limit, cooling dominates over the heating, so that the disk shrinks vertically by radiative cooling.
They found that the disk does not transit to the standard disk because the disk is supported by magnetic pressure, which prevents further vertical contraction of the disk.
\citet{dexter+2021} carried out simulations of sub-Eddington accretion flows in BHXB incorporating general relativity and radiative transfer and confirmed that the disk becomes supported by magnetic pressure.
Motivated by the numerical results of \citet{machida+2006}, \citet{oda+2009,oda+2012} obtained steady solutions of BH accretion flows supported by azimuthal magnetic fields.
They showed that the bright hard state of BHXB can be explained by the magnetically supported disk.
\citet{jiang+2019sub} and \citet{huang+2023} performed radiation MHD simulations of sub-Eddington accretion flow by numerically solving MHD equations coupled with the radiation transfer equations.
Although the accretion rate in their simulations is higher than of \citet{machida+2006}, resulting in the presence of optically thick regions, the accretion flow is still supported by the magnetic pressure \citep{jiang+2019sub,lancova+2019,dexter+2021,huang+2023}.

In this paper, we present the results of three-dimensional global radiation MHD simulations of accretion flows onto a $10^7M_{\bigodot}$ BH when the accretion rate is around $10$\% of the Eddington accretion rate.
The numerical methods and initial conditions are presented in section 2.
{
In this study, we extend the previous work on radiation MHD simulations of AGN accretion flows, as outlined by \citet{igarashi+2020}, through the inclusion of Compton scattering.
}
The numerical results are presented in section 3. 
In section 4, we compare the numerical results with steady-state solutions of magnetically supported disks.
Summary and conclusions are given in section 5.
The derivation of thermal equilibrium curves of the magnetically supported disks is summarized in the Appendix.

\section{Basic Equations and Numerical Setup} \label{sec_method}
We solve the radiation MHD equations numerically.
The basic equations in the MHD part are 
\begin{equation}
\frac{\partial\rho}{\partial t} + \bm{\nabla}\cdot\left(\rho \bm{v}\right) = 0,
\label{eqn:continuity}
\end{equation}
\begin{equation}
\frac{\partial\rho \bm{v}}{\partial t} + \bm{\nabla}\cdot\left(\rho \bm{vv} + p_{\mathrm t}\bm{I}- \frac{\bm{BB}}{4\pi} \right) =-\rho\bm{\nabla}\phi_\mathrm{PN} -\bm{S},
\label{eqn:motion}
\end{equation}
\begin{equation}
\frac{\partial {E_{\mathrm t}}}{\partial t} + \bm{\nabla}\cdot\left[\left(E_{\mathrm t}+p_{\mathrm t} \right)\bm{v} - \frac{\bm{B(v\cdot B)}}{4\pi}\right] = -\bm{\nabla}\cdot\left(\frac{1}{c}\eta\bm{j}\times\bm{B}\right)-\rho\bm{v\cdot\nabla}\phi_{\rm{PN}} - c S_\mathrm{0},
\label{eqn:energy}
\end{equation}
\begin{equation}
\frac{\partial \bm{B}}{\partial t} + \bm{\nabla\cdot}\left(\bm{vB-Bv}+\psi\bm{I}\right) = -\bm{\nabla}\times\left(\frac{4\pi}{c}\eta\bm{j}\right),
\label{eqn:induction}
\end{equation}
\begin{equation}
\frac{\partial \psi}{\partial t} + c^2_{\mathrm h}\bm{\nabla\cdot\bm{B}} = -\frac{c^2_{\mathrm h}}{c^2_{\mathrm p}}\psi,
\label{eqn:divergencefree}
\end{equation}
where $\rho$, $\bm{v}$, $\bm{B}$, and $\bm{j} = c\bm{\nabla}\times\bm{B}/4\pi$ are the mass density, velocity, magnetic field, and current density, respectively. 
In addition, $p_{\mathrm t}=p_{\rm gas}+B^2/8\pi$, and $E_{\mathrm t} = \rho v^2 /2+ p_\mathrm{gas}/(\gamma - 1) + B^2/8\pi$, where $p_\mathrm{gas}$ is the gas pressure and $\gamma=5/3$ is the specific heat ratio.
In Equations~(\ref{eqn:induction}) and (\ref{eqn:divergencefree}), $\psi$ is introduced so that the divergence-free magnetic field is maintained with minimal error during time integration, and $c_\mathrm{h}$ and $c_\mathrm{p}$ are constants \citep[see][for more details]{matsumoto+2019}.
For the gravitational potential, we adopt the pseudo-Newtonian potential to mimic general relativistic effects \citep{paczynsky+wiita1980} ;
\begin{equation}
    \phi_{\rm PN} = -\frac{GM_{\rm BH}}{R-r_{\rm s}}.
\end{equation}
Here $G$, $R=\sqrt{r^2+z^2}$, and $r_{\rm s}$ are the gravitational constant, distance to the black hole, and Schwarzschild radius, respectively.
Here, we use cylindrical coordinates  $(r,\varphi,z)$.
We assume $M_{\rm BH}=10^7M_{\bigodot}$ in all simulations. 
We adopt the so-called anomalous resistivity;
\begin{equation}
	\eta =
	\left\{
	\begin{array}{l}
	\eta_0\min\left[1,\left(v_{\mathrm d}/v_{\mathrm c}-1\right)^2\right],\ v_{\mathrm d}\geq v_{\mathrm c}, \\
	0,\ v_{\mathrm d} \leq v_{\mathrm c},
	\end{array}
	\right.
\end{equation}
where $\eta_0 = 0.01cr_{\mathrm s}$, $v_{\mathrm c}=0.9c$, and $v_{\mathrm d}=jm_{\rm p}/(\rho e)$ are the upper limits of the resistivity, critical velocity, and drift velocity, respectively. 
Here $m_{\mathrm p}$ is the proton mass, and $e$ is the elementary charge.
The resistivity $\eta$ becomes large when the drift velocity $v_{\mathrm d}$ exceeds the critical velocity $v_{\mathrm c}$ \citep[e.g.,][]{yokoyama+shibata1994}.
Equations~(1)-(5) are solved by the higher order accuracy code (5th order in space) CANS+ \citep[][and references therein]{matsumoto+2019}.
{
Note that in simulation codes based on finite volume methods, such as CANS+, numerical magnetic diffusion is unavoidable and can be larger than the anomalous resistivity.
}
In equations~(\ref{eqn:motion}) and (\ref{eqn:energy}), $\mathbf{S}$ and $S_\mathrm{0}$ are the radiation momentum and the radiation energy source terms, respectively, and are derived by solving the frequency-integrated 0th and 1st moments of the radiation transfer equations expressed in the following forms \citep[see][]{lowrie+1999,takahashi+2013,takahashi+ohsuga2013}:
\begin{equation}
\frac{\partial E_\mathrm{r}}{\partial t} + \nabla\cdot \bm{F}_\mathrm{r} = cS_\mathrm{0},
\end{equation}
\begin{equation}
\frac{1}{c^2}\frac{\partial\bm{F}_\mathrm{r}}{\partial t} + \bm{\nabla}\cdot\bm{P}_\textrm{r} = \bm{S},
\end{equation}
where
\begin{equation}
cS_\mathrm{0} = \rho\kappa_\mathrm{ff}c(a_\mathrm{r} T^{4} - E_\mathrm{r}) + \rho(\kappa_\mathrm{ff}-\kappa_\mathrm{es})\frac{\bm{v}}{c}\cdot[\bm{F}_\mathrm{r} - (\bm{v}E_\mathrm{r} + \bm{v}\cdot\mathbf{P}_\mathrm{r})] +\Gamma_{\rm c},
\end{equation}
\begin{equation}
\mathbf{S} = \rho\kappa_\mathrm{ff}\frac{\bm{v}}{c}(a_\mathrm{r}T^4 - E_\mathrm{r}) - \rho(\kappa_\mathrm{ff}+\kappa_\mathrm{es})\frac{1}{c}[\bm{F}_\mathrm{r} - (\bm{v}E_\mathrm{r} + \bm{v\cdot P}_\mathrm{r})].
\end{equation}
Here $E_{\mathrm r}, \bm{F}_{\mathrm r}, \bm{P}_{\mathrm r}$ and $a_{\mathrm r}$ are the radiation energy density, radiative flux, radiation stress tensor, and radiation constant, respectively.
In Equations (9) and (10), $\kappa_\mathrm{ff} = 1.7\times10^{-25}m_\mathrm{p}^{-2}\rho T^{-7/2}~\mathrm{cm}^2/\mathrm{g}$ is the free-free opacity and $\kappa_\mathrm{es} = 0.4~\mathrm{cm}^2/\mathrm{g}$ is the electron scattering opacity.
We solve equations~(8) and ~(9) by operator splitting.
First, we solve the equations without the source term on the right-hand side by the explicit scheme.
Then, the right-hand side (source terms) is incorporated by the implicit method \citep{takahashi+ohsuga2013}.
{
This code was applied to the radiation MHD simulations of accretion flows in CLAGN \citep{igarashi+2020} and super Eddington accretion flows around the stellar-mass black hole \citep{kobayashi+2018}.
}

We include the cooling rate $\Gamma_{\rm c}$ by the inverse Compton scattering by
\begin{equation}
	\Gamma_{\rm c} = \rho\kappa_{\rm es} cE_{\rm r0}\frac{4k_{\rm B}\left(T_{\rm e} - T_{\rm r}\right)}{m_{\rm e}c^2},
	\label{fig4:gc}
\end{equation}
where $E_{\rm r0},\ T_{\rm e}=\min{(T,10^9\ \rm{K})}, T_{\rm r}=(E_{\rm r0}/a_{\rm r})^{1/4}$, $k_{\rm B}$, and $m_{\rm e}$ are the radiation energy density in the co-moving frame, the electron temperature, the radiation temperature, Boltzmann constant, and electron mass, respectively \citep[e.g.,][]{kawashima+2009,padmanabhan2000}.
For simplicity, we solve the equations for the single-temperature plasma while partially accounting for the effects of the two-temperature plasma (i.e., the plasma in which electron temperature differs from the ion temperature).
The two-temperature plasmas appear in the corona and accretion flows with accretion rates well below the Eddington accretion rate.
Since the ion cooling rate through collisions with electrons is low in nearly collisionless hot plasmas, ions remain hot meanwhile electrons can be cooled by radiation.
Two temperature models of RIAFs have been extensively studied \citep[e.g.,][]{nakamura+1997}.
{
Electron temperature in general relativistic radiation MHD simulations of two-temperature accretion flows is $T_{\rm e}=10^9-10^{10}\ \rm{K}$ in RIAFs \citep[e.g.,][]{sadowski+2017,ryan+2017,ryan+2018} and $T_{\rm e}\sim10^9\ \rm{K}$ when the accretion rate exceeds $10^{-3}\dot{M}_{\rm Edd}$ \citep{dexter+2021,liska+2022a}.
}
%

%
For the studies of hot accretion flows with relatively higher mass accretion rates, the single-temperature plasma approximation will lead to overcooling of the ions  \citep[see e.g., general relativistic radiation MHD simulation with Monte Carlo radiative transfer by][]{ryan+2015}.
To avoid the overcooling of ions, we introduce an upper limit for the electron temperature, because inverse Compton scattering in hot accretion flows cools the electrons down to $10^9$ K.

In the low-temperature region, a floor value for the gas temperature $T=T_{\rm floor}=5\times10^5\ \rm{K}$ is introduced to avoid negative gas pressure when the gas pressure is derived from the total energy which includes the thermal energy, the magnetic energy, and the kinetic energy.
Note that the total energy is used as a conserved variable.

{
In simulations presented in this paper, we first perform a non-radiative simulation starting from a rotating equilibrium torus embedded in a weak poloidal magnetic field with initial $\rm{plasma}\ \beta=10$ until a quasi-steady BH accretion flow is formed.
To describe a pure poloidal initial magnetic field, we set a $\varphi$ component of the vector potential which is proportional to the density ($A_\varphi\propto\rho$) of the initial torus \citep[see][]{hawley+balbus2002,kato+2004}.
}

The computational domain of our simulation is $0 \le r < 2000r_{\rm s}$, $0 \le \varphi < 2\pi$, and $|z| < 2000r_{\rm s}$, and the number of grid points is $(n_{r},n_\varphi,n_{z}) = (464,32,464)$. 
The grid spacing is $0.1r_{\rm s}$  in the radial and vertical directions when $r < 20r_{\rm s}$ and $|z| < 5r_{\rm s}$ and increases outside the region.
The absorbing boundary condition is imposed at $R=2r_{\rm s}$, and the outer boundaries are free boundaries where waves can transmit.
%

\begin{figure}[htbp]
 	\centering
    \plotone{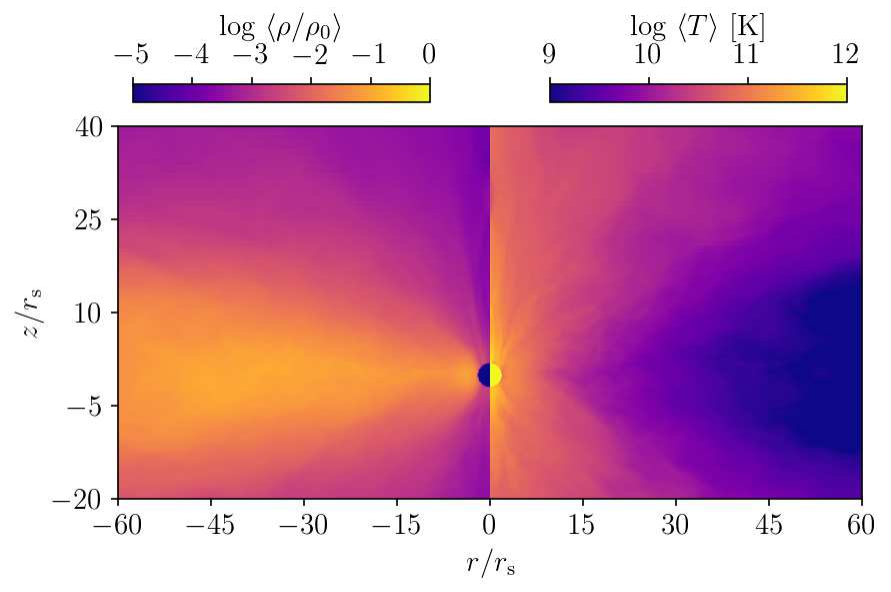}
	\caption{Azimuthally averaged distribution of the gas density (left) and gas temperature (right) in the $r-z$ plane when the radiative cooling term is turned on ($t/(10^4t_0)=1.05$).}
	\label{fig1}
\end{figure}
\begin{figure}[htbp]
 	\centering
	\plotone{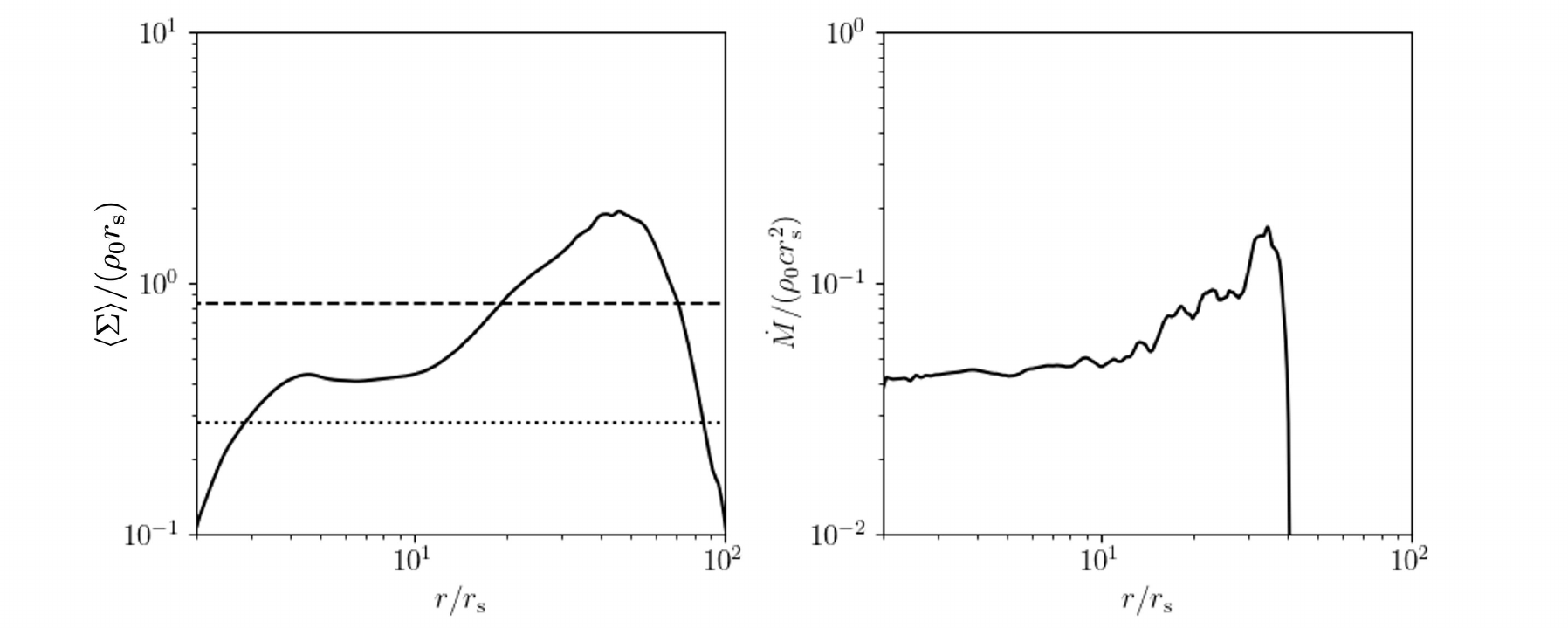}
	\caption{
			{Radial distribution of the azimuthally averaged surface density (left) and the net mass accretion rate (right) averaged over $0.85<t/(10^4t_0)<1.05$.
			The dashed and dotted lines in the left panel show $\tau_{\rm es}=10$ for model M01 and model M03, respectively}}
	\label{fig2}
\end{figure}

%
Figure~\ref{fig1} shows the azimuthally averaged distribution of the density $\langle\rho\rangle$ (left) and temperature $\langle T \rangle$ (right) at  $t/(10^4t_0)=1.05$.
{
Here, $\displaystyle t_0=r_{\mathrm s}/c=100\ \left( \frac{M_{\rm BH}}{10^7M_{\bigodot}} \right) \rm{s}$ and $\rho_0$ is the maximum density of the initial torus.
In this paper, $\langle A \rangle$ denotes the azimuthal average calculated as 
\begin{equation}
	\langle A \rangle = \frac{\int_0^{2\pi} A d\varphi}{\int_0^{2\pi} d\varphi},
\end{equation}
$\langle A \rangle_\rho$ denotes the density-weighted azimuthal average calculated as 
\begin{equation}
	\langle A \rangle_{\rho} = \frac{\int_0^{2\pi} A\rho d\varphi}{\int_0^{2\pi} \rho d\varphi},
\end{equation}
and $\langle A \rangle_{\rm{V}}$ denotes the volume average calculated as 
\begin{equation}
	\langle A \rangle_{V} = \frac{\iiint A drrd\varphi dz}{\iiint drrd\varphi dz}.
\end{equation}
}
Since the radiative cooling term is not included until $t/(10^4t_0)=1.05$, the accretion flow is hot ($T>10^8$ K) throughout the region.
The radiative cooling term is turned on at this stage.
{
The structure of the non-radiative accretion flow produced by this simulation is described in \citet{igarashi+2020}.
}
The left panel of Figure~\ref{fig2} shows the radial distribution of the azimuthally averaged surface density 
\begin{equation}
	\Sigma = \int^{100r_{\rm s}}_{-100r_{\rm s}} \rho dz
\end{equation} 
{averaged before including the radiation term ($0.85<t/(10^4t_0)<1.05$).}
The surface density is highest around $r=40r_{\rm s}$.
The density enhancement is the remnant of the initial torus.
The right panel of Figure~\ref{fig2} shows the net mass accretion rate computed by;
\begin{equation}
    \dot{M} = -\int^{100r_{\rm s}}_{-100r_{\rm s}}\int^{2\pi}_0 \rho v_r rd\varphi dz,
\end{equation}
where $v_r$ is the radial velocity in cylindrical coordinates.
The net mass accretion rate is constant up to {$\sim20r_{\rm s}$}.
The unit density $\rho_0$ is {set so} that this accretion rate {is} equal to the accretion rate specified as a parameter.

In this paper, numerical results for model M01 ($\dot{M}\sim0.1\dot{M}_{\rm{Edd}}$, $\rho_0=1.0\times10^{-11}\ \rm{g/cm^3}$), and model M03 ($\dot{M}\sim0.3\dot{M}_{\mathrm{Edd}}$, $\rho_0=3.0\times10^{-11}\ \rm{g/cm^3}$) are presented.
Here, $\dot{M}_{\mathrm{Edd}}$ is the Eddington accretion rate defined by $\dot{M}_{\mathrm{Edd}}=L_{\mathrm{Edd}}/c^2$, where $L_{\mathrm{Edd}}$ is the Eddington luminosity.
Table~\ref{tab:norm} summarizes the normalizations and parameters used in this paper.
\begin{table}[htbp]
    \centering
    \begin{tabular}{ccc}
        \hline
        \hline
        Black hole mass & $\displaystyle M_{\rm BH}=10^7M_{\bigodot}$ \\
        Velocity & $\displaystyle c=3\times10^{10}\ \rm{cm/s}$ \\
        Length & $\displaystyle r_{\rm s}=3\times10^{12}\left(\frac{M_{\rm BH}}{10^7M_{\bigodot}}\right)\ \rm{cm}$ \\
        Eddington Luminosity & $\displaystyle L_{\rm Edd}=\frac{4\pi cGM_{\rm BH}}{\kappa_{\rm es}}\ \rm{erg/cm^2}$ \\
        Eddington accretion rate & $\displaystyle \dot{M}_{\rm Edd} = L_{\rm Edd}/c^2 $ \\
    \end{tabular}
     \centering
     \begin{tabular}{ccc}
        \hline
        \hline
          & M01 & M03 \\
        \hline
        Density & $\displaystyle \rho_0=1\times 10^{-11}\ \rm{g/cm^3}$ & $\rho_0=3\times 10^{-11}\ \rm{g/cm^3}$ \\
        Energy and Pressure& $\displaystyle E_0 = P_0 =  \rho_0c^2=9\times10^{9}\ \rm{erg/cm^3}$ &  $E_0 = P_0 =  \rho_0c^2=2.7\times10^{10}\ \rm{erg/cm^3}$\\
        Flux & $\displaystyle F_0 = \rho_0c^3=2.7\times10^{20}\ \rm{erg/cm^2/s}$  & $F_0 = \rho_0c^3=5.1\times10^{20}\ \rm{erg/cm^2/s}$ \\
        Magnetic field & $\displaystyle B_0 = \sqrt{4\pi\rho_0 c^2}\sim 3.7\times10^5\ \rm{G}$ &$\displaystyle B_0 = \sqrt{4\pi\rho_0 c^2}\sim 6.4\times10^5\ \rm{G}$ \\
        \hline
    \end{tabular}
    \caption{{Normalizations and parameters used in this paper.}}
    \label{tab:norm}
\end{table}%

%
The horizontal lines in the left panel of Figure~\ref{fig2} show the surface density corresponding to $\tau_{\rm es}(0)=(\tau_{\rm es}(+0)+\tau_{\rm es}(-0))/2=10$ for Model M01 and M03, where $\tau_{\rm es}(z)$ is the electron scattering optical depth calculated by
\begin{equation}
 \tau_{\rm es}\left( z \right) = 
 \ \left\{ 
 \begin{array}{ll}
 	 \int^{100r_{\rm s}}_z\rho\kappa_{\rm es}dz,\ {\rm when}\ z>0\\
	 \int^z_{-100r_{\rm s}}\rho\kappa_{\rm es}dz,\ {\rm when}\ z<0
 \end{array}
 \right.
 \end{equation}
For model M01, $\tau_{\rm es}(0)$ exceeds $10$ in the region $20r_{\rm s}<r<80r_{\rm s}$.
In model M03, $\tau_{\rm es}(0)$ exceeds $10$ in $3r_{\rm s}<r<80r_{\rm s}$.

%
\section{Numerical results}
\subsection{Formation of the Soft X-ray Emitting, Warm Comptonization Region}
\begin{figure}[htbp]
\centering
\plotone{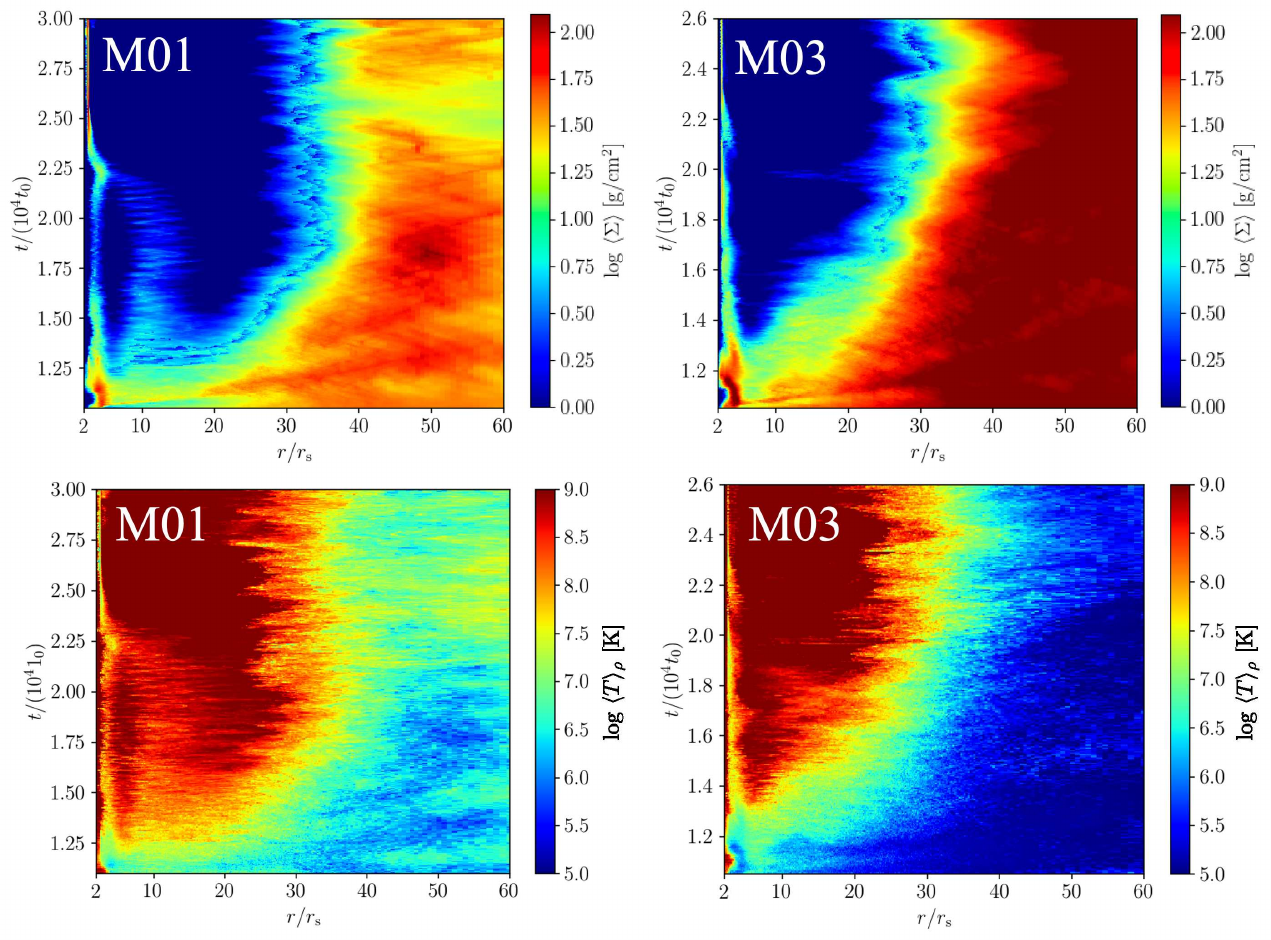}
\caption{{Time evolution of azimuthally averaged surface density (upper panels) and azimuthally averaged density-weighted equatorial temperature (lower panels) for model M01 (left panels) and M03 (right panels).}}
\label{fig3}
\end{figure}

{
Figure~\ref{fig3} shows the time evolution of the azimuthally averaged surface density $\langle\Sigma\rangle$ (upper panels) and the density-weighted azimuthally averaged equatorial temperature $\langle T \rangle_\rho$ (lower panels) for model M01 (left panels) and M03 (right panels).
After the inclusion of the radiative cooling term at $t/(10^4t_0)=1.05$, the accretion flow quickly cools and warm disk with temperature $T=10^6-10^{7.5}\ \rm{K}$ is formed at $r>35r_{\rm s}$ in M01 and cool dense disk with temperature $T=10^5-10^6\ \rm{K}$ is formed at $r>40r_{\rm s}$ in M03.
}
%

\begin{figure}[htbp]
\centering
\plotone{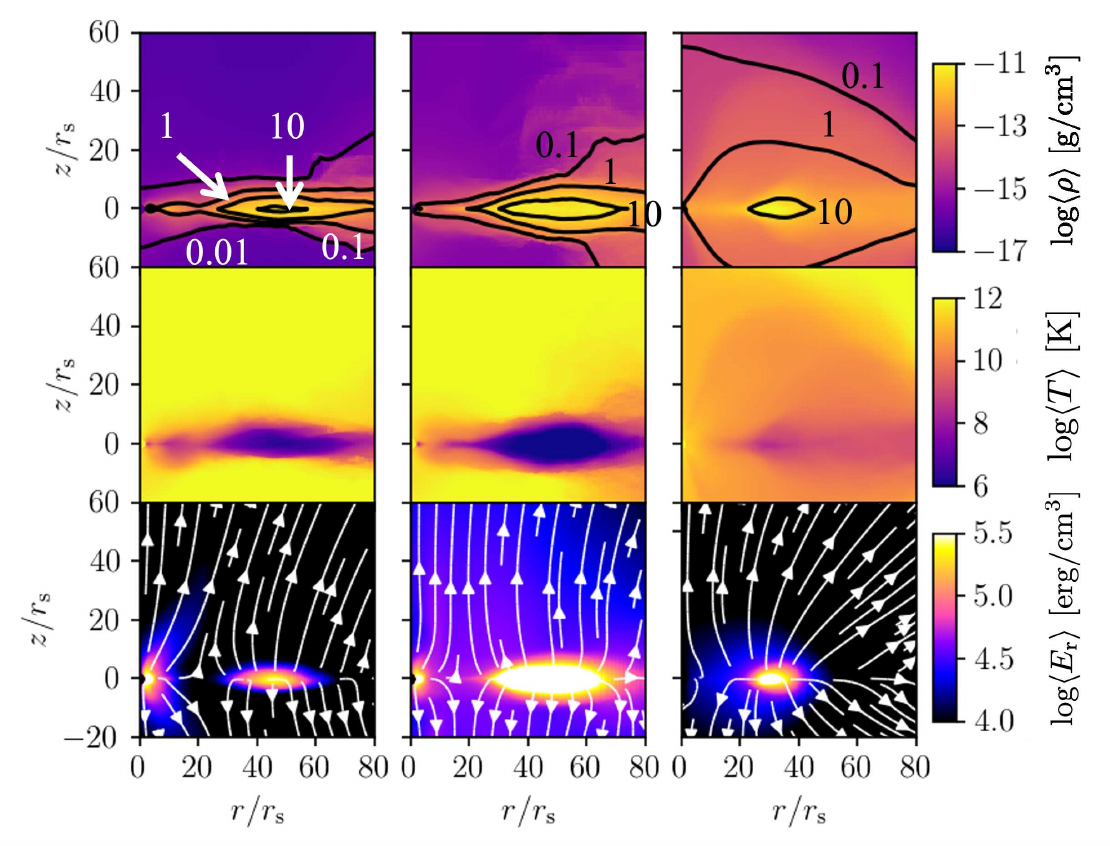}
\caption{Distribution of the azimuthally averaged gas density (top), gas temperature (middle), and radiation energy density (bottom) in the poloidal plane.
		The left panels show the results for model M01 averaged over $1.5<t/(10^4t_0)<1.75$, and the middle panels show the results for model M03 averaged over $1.55<t/(10^4t_0)<1.8$. 
		The contours in the top panels show the location where $\tau_{\mathrm{es}}(z)=0.01, 0.1, 1, 10$.
		The white streamlines in the bottom panels show the radiation energy flux in the poloidal plane.
		The right panel shows the results of the model without inverse Comptonization (model NC), in which the accretion rate is 10\% of the Eddington accretion rate.
		The quantities in model NC are averaged over the time range $3.6<t/(10^4t_0)<3.8$.}
\label{fig4}
\end{figure}

%
Figure~\ref{fig4} shows the azimuthally averaged density $\langle\rho\rangle$ (top), temperature $\langle T \rangle$ (middle), and radiation energy density $\langle E_{\rm r} \rangle$ (bottom) distributions when the warm disk is formed.
The left panels are for model M01 averaged over $1.5<t/(10^4t_0)<1.75$, and the middle panels are for model M03 averaged over $1.55<t/(10^4t_0)<1.8$. 
The right panel shows results without Compton cooling/heating reported by \citet{igarashi+2020} (model NC, $\dot{M}\sim0.1\dot{M}_{\rm Edd}$, $\rho_0=2.0\times10^{-11}\ \rm{g/cm^3}$).
The contours in the top panels of Figure~\ref{fig4} show isocontours of $\tau_{\rm es}(z)$. 
In the model with Comptonization, the disk thickness and temperature decrease due to Compton cooling.
In the outer region ($r>20r_{\rm s}$), the gas temperature decreases significantly to $\sim10^7$ K in model M01 and $\sim10^6$ K in model M03, and a warm, Thomson-thick region is formed.
In models M01 and M03, the radiation energy density in the warm region is large.
In the region where $\tau_{\rm es}(z)>10$, radiation is trapped in this region and the region is radially wider than in the model without Comptonization. 
The radiation energy density in model M01 is comparable to that in model NC.
On the other hand, model M03 has a higher radiation energy density than the other models due to the larger optical depth in the warm region. 
{
The white streamlines in the bottom panels show the direction of the radiative flux in the poloidal plane.
The direction of the radiative flux is almost vertical in the region where $\tau_{\rm es}<1$.
}

%

{
The bolometric luminosity $L=\int F_z(r,\varphi,z=40r_{\rm s})drrd\varphi$ of Model M01 is $L\sim0.01L_{\rm Edd}$, consistent with the X-ray intensity at the onset of the changing look state transition.
The luminosity increases with increasing the mass accretion rate, reaching $L\sim0.08L_{\rm Edd}$ in model M03.
This luminosity is consistent with that in the UV and soft X-ray dominant state of CLAGN.
}

%

\begin{figure}[htbp]
 	\centering
	\plotone{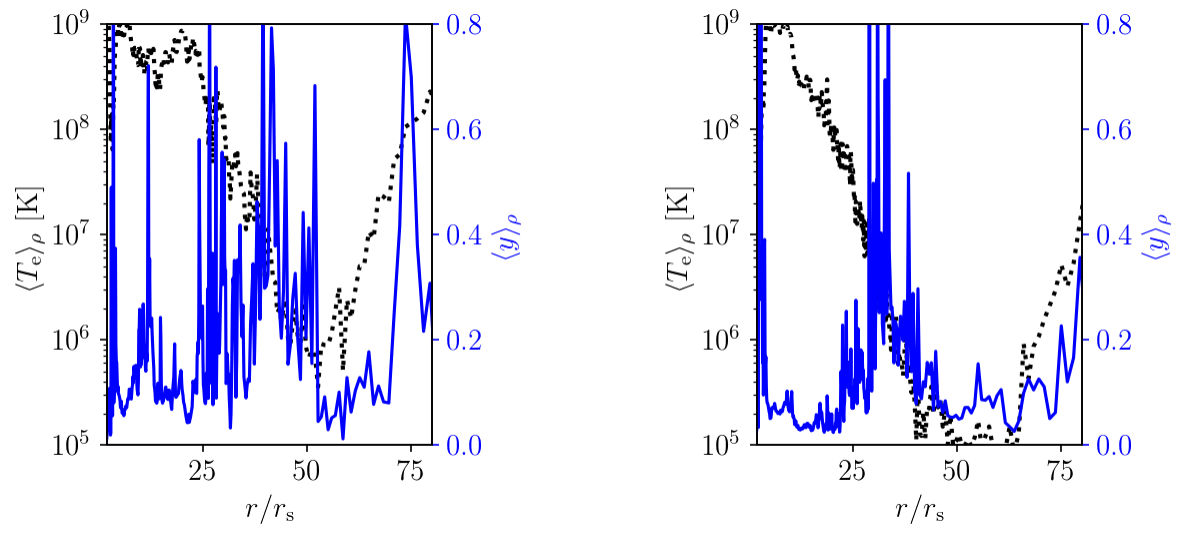}
	\caption{The radial distribution of the {density-weighted azimuthally averaged electron temperature (black dotted line)} and Compton $y$-parameter (blue solid line) at mid-plane for model M01 (left) and model M03 (right).
The time of the snapshot is $t=1.75\times10^4t_0$. }
	\label{fig5}
\end{figure}

%
To identify the soft X-ray emitting region, we calculate the Compton $y$-parameter defined by
\begin{equation}
	y = \frac{4k_{\rm B}T_{\rm e}}{m_{\rm e}c^2}\max{(\tau_{\rm es}(0), \tau_{\rm es}(0)^2)}.
\end{equation}
When $T>10^9\ \rm{K}$, $T_{\rm e}$ is set equal to $10^9\ \rm{K}$, as discussed in the previous section.
As we also mentioned in Section 2, we set the lower limit of temperature as $5\times10^5\ \rm{K}$ to avoid the negative gas pressure in the simulation.
This can lead to the overestimation of the Compton $y$-parameter.
To avoid this, we replace the electron temperature $T_{\rm e}$ with radiation temperature $T_{\rm r}$ in the region where $T<10^6\ \rm{K}$ when we compute the Compton $y$-parameter.
Since $T_{\rm r}$ can be less than $5\times10^5\ \rm{K}$, our estimate gives a lower limit of the Compton $y$-parameter.
Figure~\ref{fig5} shows the radial distribution of {density-weighted azimuthally averaged $T_{\rm e}$ (black dotted curves)} and the Compton $y$-parameter (blue solid curves).
In model M01, the warm region with a temperature $T=10^6-10^7\ \rm{K}$ lies outside $\sim40r_{\rm s}$.
The Compton $y$-parameter in this region is $0.3-0.4$.
The CLAGN {observation} suggests that the Compton $y$-parameter of the soft X-ray emitting region is $\sim0.4$ and less than $1$ \citep{noda+done2018,tripathi+dewangan2022}.
This is consistent with our simulation results.
The Compton $y$-parameter for M01 is $0.1-0.2$ in the hot inner region ($r<30r_{\rm s}$) where $T_{\rm e}\sim10^9$ K.
The hard X-rays can be emitted from this region by inverse Compton scattering of soft photons.
In M03, the warm region ($T=10^6-10^7\ \rm{K}$) moves inward and is located around $30r_{\rm s}$.
The Compton $y$-parameter in this warm region is $0.3\sim0.4$, so we expect soft X-ray emission from this region.
In the outer region where $r>40r_{\rm s}$, since the electron temperature drops to $10^5$ K, we expect UV emission from this region.
%

%
Table~\ref{tab:feature} summarizes the observational features of our simulations.
In model M01, the soft X-ray emitting region appears around $r=40r_{\rm s}$ but the UV emitting region does not appear because the temperature exceeds $10^6$ K.
In model M03, temperature in $r>50r_{\rm s}$ decreases down to $10^5$ K (see figure~\ref{fig5}) and emits UV radiation.
In model NC, the Compton $y$-parameter is larger but the temperature exceeds $10^7$ K because Compton cooling is not included \citep{igarashi+2020}.
\begin{table}[htbp]
    \centering
    \begin{tabular}{ccccc}
        \hline
        Model & Disk Luminosity & Soft X-ray emitting region & UV emitting region &  Compton-$y$ in the warm region\\
        \hline
        \hline
        M01 & $\sim0.01L_{\rm Edd}$ & Yes & No & $\sim0.3$  \\
        M03 & $\sim0.08L_{\rm Edd}$ & Yes & Yes & $\sim0.4$   \\
        NC & $\sim0.005L_{\rm Edd}$ & Yes & No & $\sim1$ \\
        \hline
    \end{tabular}
    \caption{Observational features of the simulations.
    			Note that the Thomson optical depth and temperature of the soft X-ray emitting region is $\tau_{\rm es}\sim10$ and $T=10^6-10^7\ \rm{K}$. The temperature of the UV emitting region is $T\sim10^5\ \rm{K}$.}
    \label{tab:feature}
\end{table}%

\begin{figure}[htbp]
\centering
\plotone{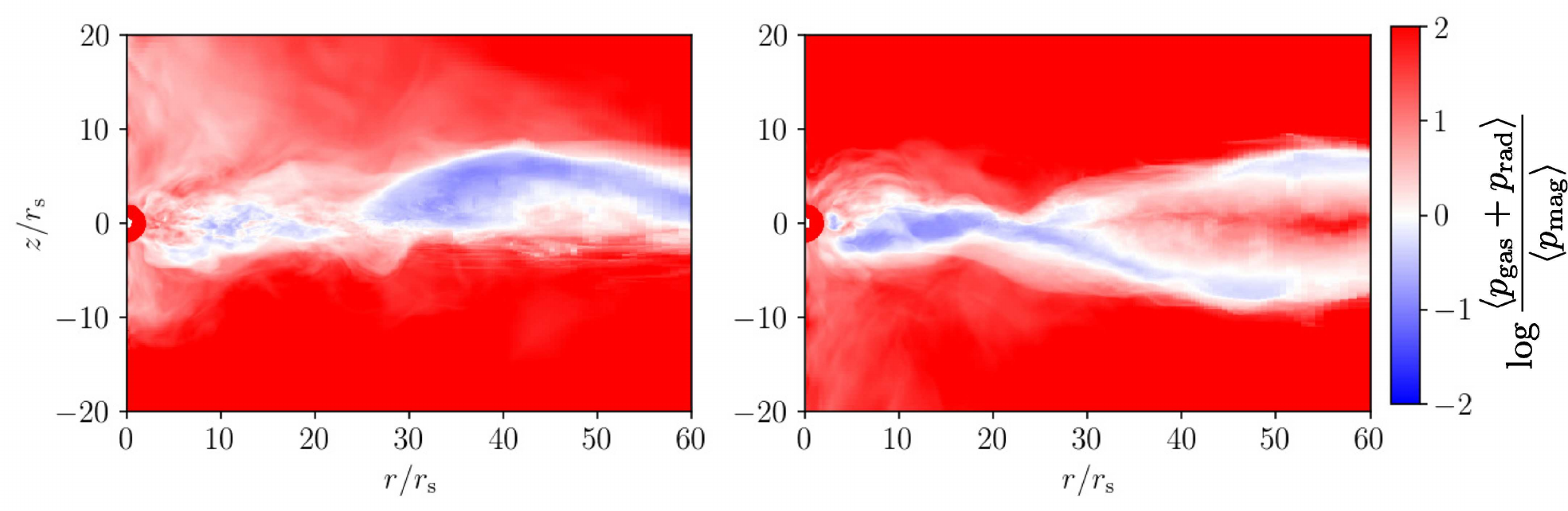}
\caption{Distribution of the $\langle p_{\rm gas}+p_{\rm rad} \rangle/\langle p_{\rm mag} \rangle$ at $t=1.75\times10^4t_0$ for model M01 (left) and model M03 (right).}
\label{fig6}
\end{figure}

%
Figure~\ref{fig6} shows the distribution of $\langle p_{\rm gas}+p_{\rm rad}\rangle/\langle p_{\rm mag}\rangle$ for model M01 (left) and M03 (right), {where $p_{\rm rad}=E_{\rm r}/3$ is the radiation pressure.}
In M01, the magnetic pressure is dominant in the warm region in the upper half-hemisphere and in the hot inner torus.
In model M03, the magnetic pressure dominates in the inner torus and in the warm region ($20r_{\rm s}<r<30r_{\rm s}$).
The radiation pressure dominates in the outer region $(r>30r_{\rm s}).$
For M01, the accretion rate is around $10$\% of the Eddington accretion rate, and the radiation pressure is lower than that in model NC (without Comptonization) reported in \citet{igarashi+2020}.
This is because the disk temperature in M01 is lower than that in NC due to Compton cooling.
%

\begin{figure}[htbp]
\centering
\plotone{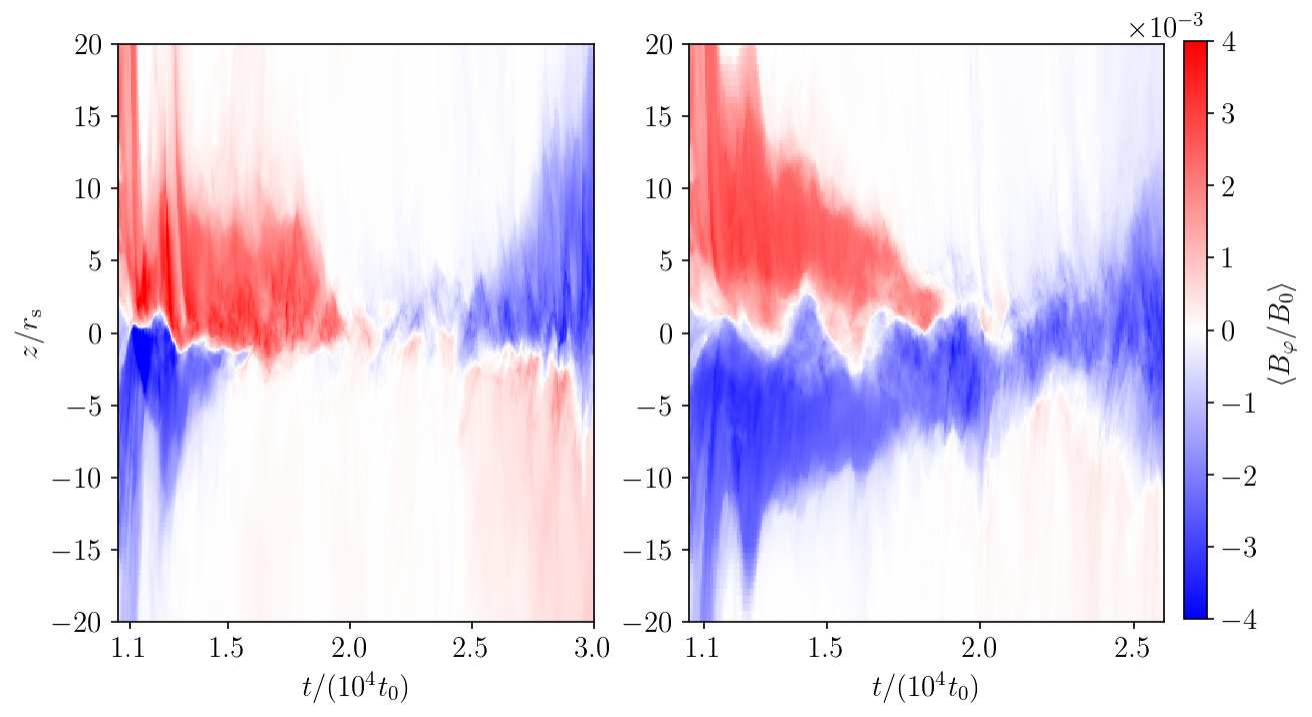}
\caption{The distribution of the azimuthally averaged azimuthal magnetic field at $r=40r_{\rm s}$ for model M01 (left) and M03 (right).
		Red and blue show different polarity.
		The horizontal axis shows the time and the vertical axis shows the height.}
\label{fig7}
\end{figure}

%
In model M01, the magnetic pressure dominant warm region appears only above the equatorial plane because the azimuthal magnetic fields below the equatorial plane reconnect with those above the equatorial plane.
Figure~\ref{fig7} shows the butterfly diagram of the azimuthally averaged azimuthal magnetic fields at $r=40r_{\rm s}$.
During the vertical contraction of the disk due to radiative cooling, azimuthal magnetic fields anti-symmetric to the equatorial plane reconnect, and in model M01, only positive (red) magnetic fields remain in the upper hemisphere.
This is the reason why the magnetic pressure dominant region in figure~\ref{fig6} appears only in the upper hemisphere around $r=40r_{\rm s}$.
In the later stage ($t/(10^4t_0)>2.0$), the azimuthal magnetic fields are again amplified by magnetorotational instability, and their polarity reverses.
In model M03, azimuthal magnetic fields remain both in the lower and upper hemispheres at $t/(10^4t_0)=1.75$.
%

%

\subsection{Heating by the Magnetic Reconnection around the Equatorial Plane}
\begin{figure}[htbp]
\centering
\plotone{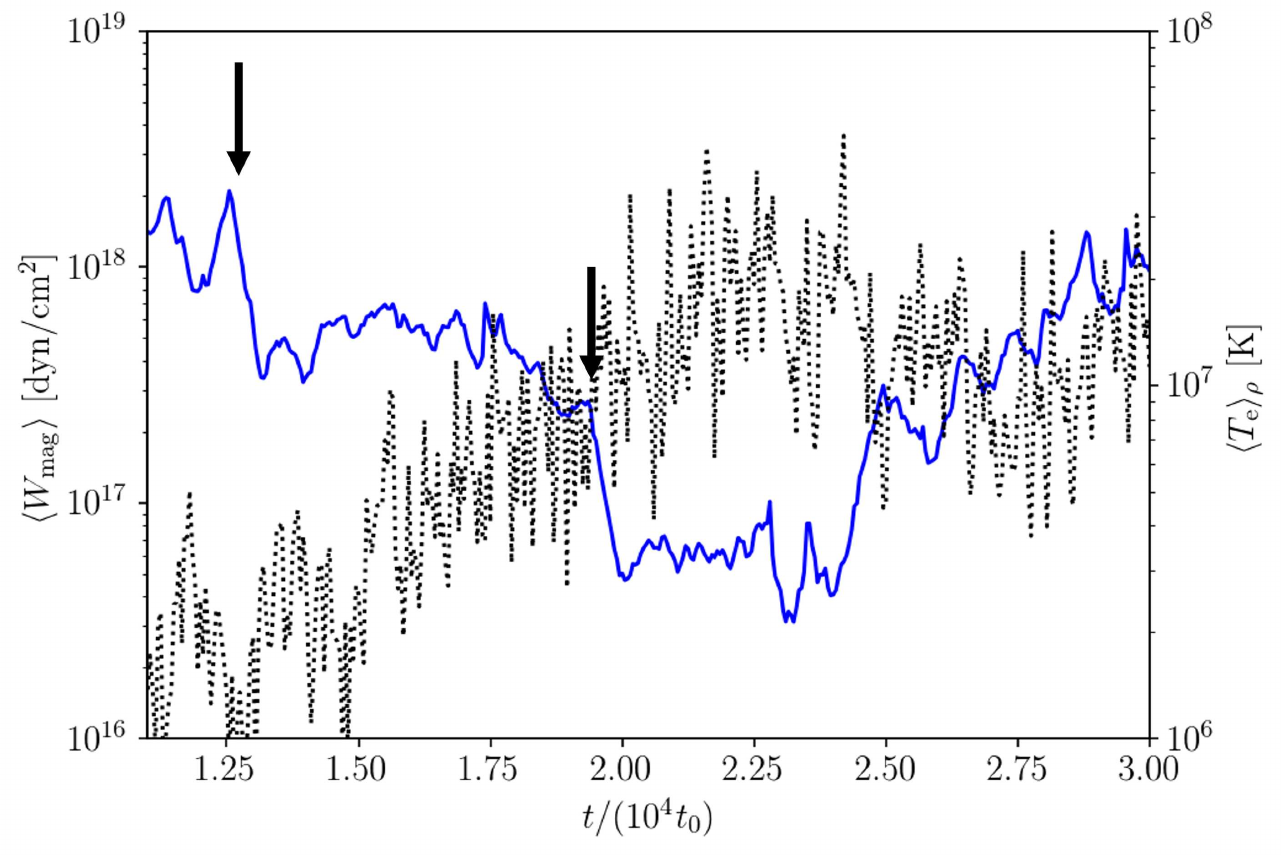}
\caption{The time evolution of the azimuthally averaged vertically integrated magnetic pressure (blue solid line) and the {density-weighted azimuthally averaged equatorial electron temperature (black dashed line)} at $r=40r_{\rm s}$ for model M01.
		Arrows indicate the time when magnetic energy is released.}
\label{fig8}
\end{figure}

{
Figure~\ref{fig8} shows the time variation of the azimuthally averaged vertically integrated magnetic pressure (blue solid line), and the density-weighted azimuthally averaged equatorial temperature (black dashed line) for model M01 at $r=40r_{\rm s}$.
Around $t/(10^4t_0)\sim1.3$ and $t/(10^4t_0)\sim1.9$, the magnetic pressure decreases, and the equatorial temperature increases.
It indicates that magnetic energy is converted to thermal energy.
}
%

\begin{figure}[htbp]
\centering
\plotone{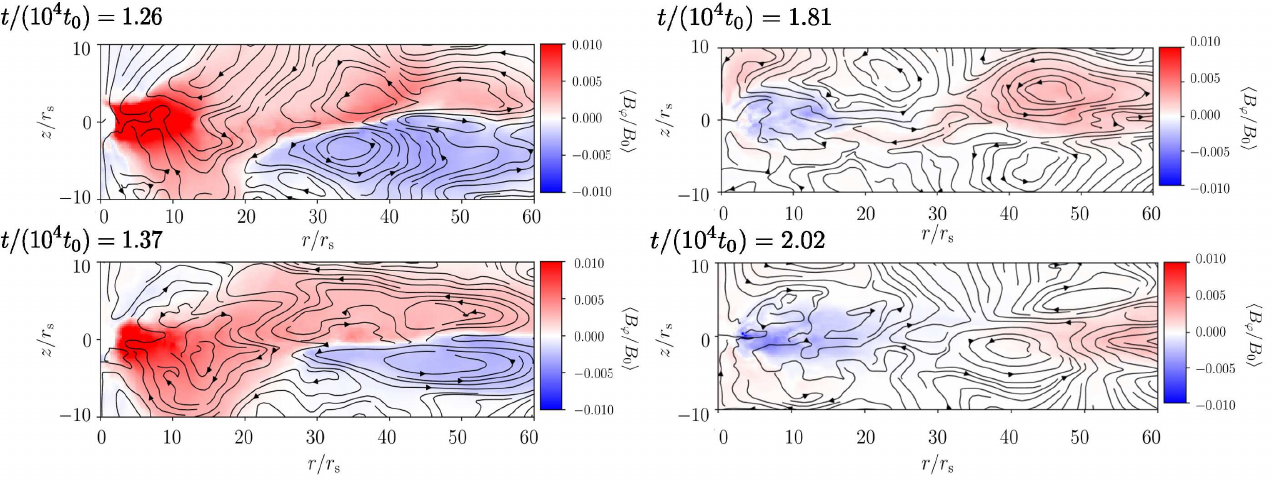}
\caption{
	The azimuthally averaged azimuthal magnetic field (color) and magnetic field lines in the poloidal plane (black streamlines) at $t/(10^4t_0)=1.26$ (top left panel) and $t/(10^4t_0)=1.37$ (lower left panel) and $t/(10^4t_0)=1.81$ (top right panel) and $t/(10^4t_0)=2.02$ (lower right panel)}
\label{fig9}
\end{figure}

Figure~\ref{fig9} shows the azimuthally averaged azimuthal magnetic field and poloidal magnetic field lines before and after $t/(10^4t_0)\sim1.3$ (left panels) and $t/(10^4t_0)\sim1.9$ (right panels).
{
In the left panels, azimuthal magnetic flux tubes with opposite poloidal and azimuthal magnetic fields are moving outward.
These helical flux tubes are formed in the inner region ($r\sim20r_{\rm s}$) at around $t/(10^4t_0)=1.2$.
They collide at the equatorial plane and merge.
This event is similar to the merging of two spheromacs with opposite current helicity $\bm{j\cdot B}$.
The merging of the counter-helicity flux tubes converts the magnetic energy into thermal energy and heats the plasma \citep{ono+1996}. 
The right panels of figure~\ref{fig9} show that the azimuthal magnetic field in the inner hot flow ($r<20r_{\rm s}$) is reversed from that in the left panels and magnetic reconnection with that in the outer warm disk takes place.
}

{
The outward motion of the helical flux tubes produces the outward motion of the plasma in the equatorial region.
Figure~\ref{fig10} shows the space-time plot of the accretion rate at $|z|>3r_{\rm s}$ (left panel), $|z|<3r_{\rm s}$ (middle panel), and the net mass accretion rate (right panel) for M01.
Before the radiative cooling is included ($t/(10^4t_0)<1.05$), accretion proceeds in the equatorial region of the disk but as the disk cools, accretion takes place mainly in the surface region of the disk, and outward motion becomes prominent in the equatorial region.
It indicates that meridional circulation is taking place in the radiatively cooled warm disk at $r>30r_{\rm s}$.
The surface accretion flow is driven by the angular momentum loss of the plasma around the disk surface through large-scale poloidal magnetic fields threading the disk.
The surface accretion flow was called 'avalanche flow' in \citet{matsumoto+1996}.
\begin{figure}[htbp]
\centering
\plotone{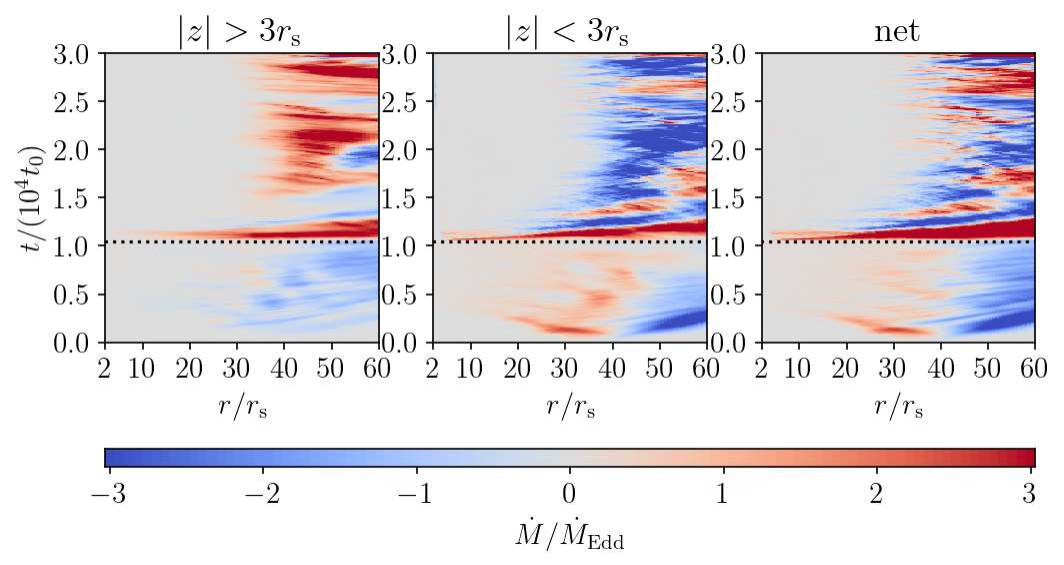}
\caption{{The time evolution of the mass accretion rate in $r-t$ plane for model M01.
		The left, middle, and right panels show the mass accretion rate in the surface region ($|z|>3r_{\rm s}$), equatorial region ($|z|<3r_{\rm s}$), and the net mass accretion rate.
		The blue and red regions represent the outflow and inflow regions.}}
\label{fig10}
\end{figure}
}

{
Figure~\ref{figadd2} shows the distribution of the Poynting flux $\displaystyle \bm{F}_{\rm poy} = -\bm{(v\times B)\times B}/4\pi$ at $t/(10^4t_0)=1.25$ (left), and $t/(10^4t_0)=1.5$ (right). 
In the warm region at $r>35r_{\rm s}$, radially outward Poynting flux is comparable to the radiative flux in this region.
It indicates that the magnetic energy transported to this region contributes to the heating of the plasma.
Heating of the accretion flow by Poynting flux from the inner region has been studied by general relativistic radiation MHD simulations by \citet{takahashi+2016}.
Our numerical results indicate that the Poynting flux contributes to the heating of the warm Compton region of the disk even when the central black hole is not rotating.
\begin{figure}[htbp]
\centering
\plotone{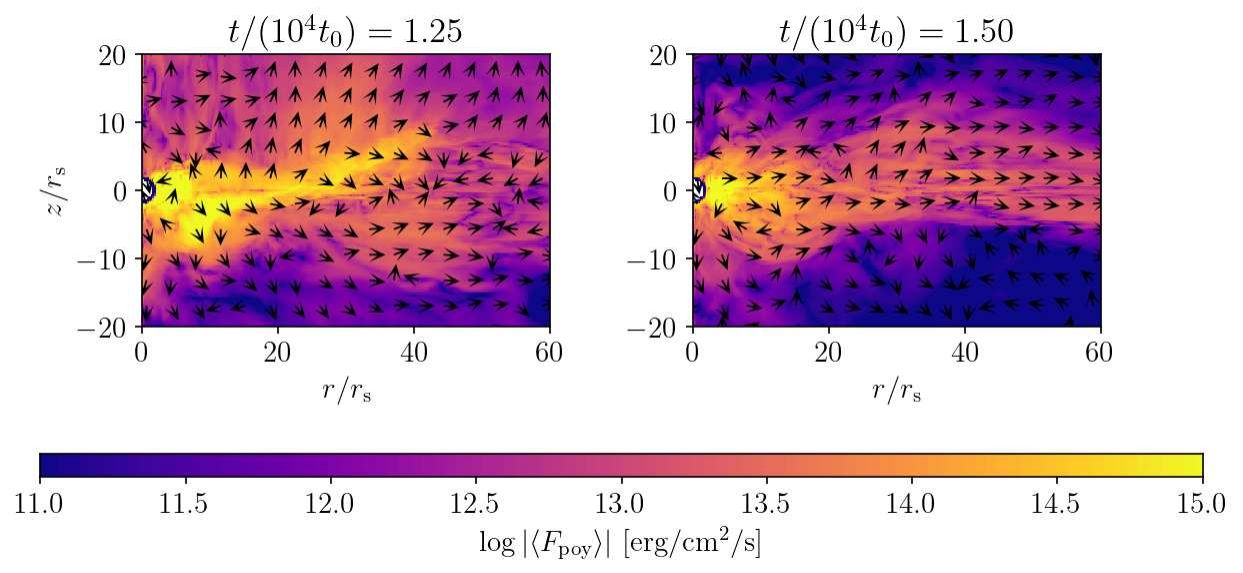}
\caption{{The absolute value of the azimuthally averaged poloidal Poynting flux at $t/(10^4t_0)=1.25$ (left) and $t/(10^4t_0)=1.5$ (right) for model M01.
		 The black arrows show the direction of the flux in the poloidal plane.}}
\label{figadd2}
\end{figure}
}

\section{Comparison with Models of Magnetized Disks}
We showed by radiation MHD simulations including Compton cooling/heating that the warm Comptonization region supported by azimuthal magnetic fields is formed in the region outside the radiatively inefficient accretion flow near the black hole.
In this region, the disk shrinks in the vertical direction by radiative cooling.
In this section, we compare numerical results with steady models of magnetized black hole accretion flows.
The basic equations of steady accretion disks partially supported by the magnetic pressure of the azimuthal magnetic field were derived by \citet{oda+2009} assuming that the heating term in the energy equation is computed by using $Q^+=-\alpha W_{\rm tot}rd\Omega/dr$ where $\Omega$ is the angular speed of the rotating disk, and $W_{\rm tot}$ is the vertically integrated total pressure $p_{\rm tot}=p_{\rm gas}+p_{\rm rad}+p_{\rm mag}$.
Here we update their model by including additional magnetic heating due to magnetic reconnection and assume that $Q^+=-\alpha W_{\rm tot}rd\Omega/dr+(3/2)\alpha'W_{\rm tot}\Omega$.
The derivation of thermal equilibrium curves for this model is shown in Appendix.
In this section, we compare the numerical results with the magnetized disk model.

%
{
Here, we examine the angular momentum transport rate, $\alpha$ which is calculated by the following equation.
\begin{equation}
	\alpha = \frac{\langle -B_rB_\varphi/4\pi\rangle_V}{\langle p_{\rm tot} \rangle_V}.
\end{equation}
Figure~\ref{fig11} shows the time evolution of $\alpha$ averaged in  $35r_{\rm s}<r<45r_{\rm s}$.
The dashed curve shows $\alpha$ in the equatorial region where $\tau_{\rm es}>1$, and the solid curve shows $\alpha$ in the surface region where $0.1<\tau_{\rm es}<1$.
The $\alpha$ value calculated in the surface region (solid curve) is larger than that of the disk region.
The smaller $\alpha$ in the equatorial region is due to the inhibition of accretion through the radially outward motion of helical flux tubes.
In the following, we adopt $\alpha=0.1$.
\begin{figure}[htbp]
\centering
\plotone{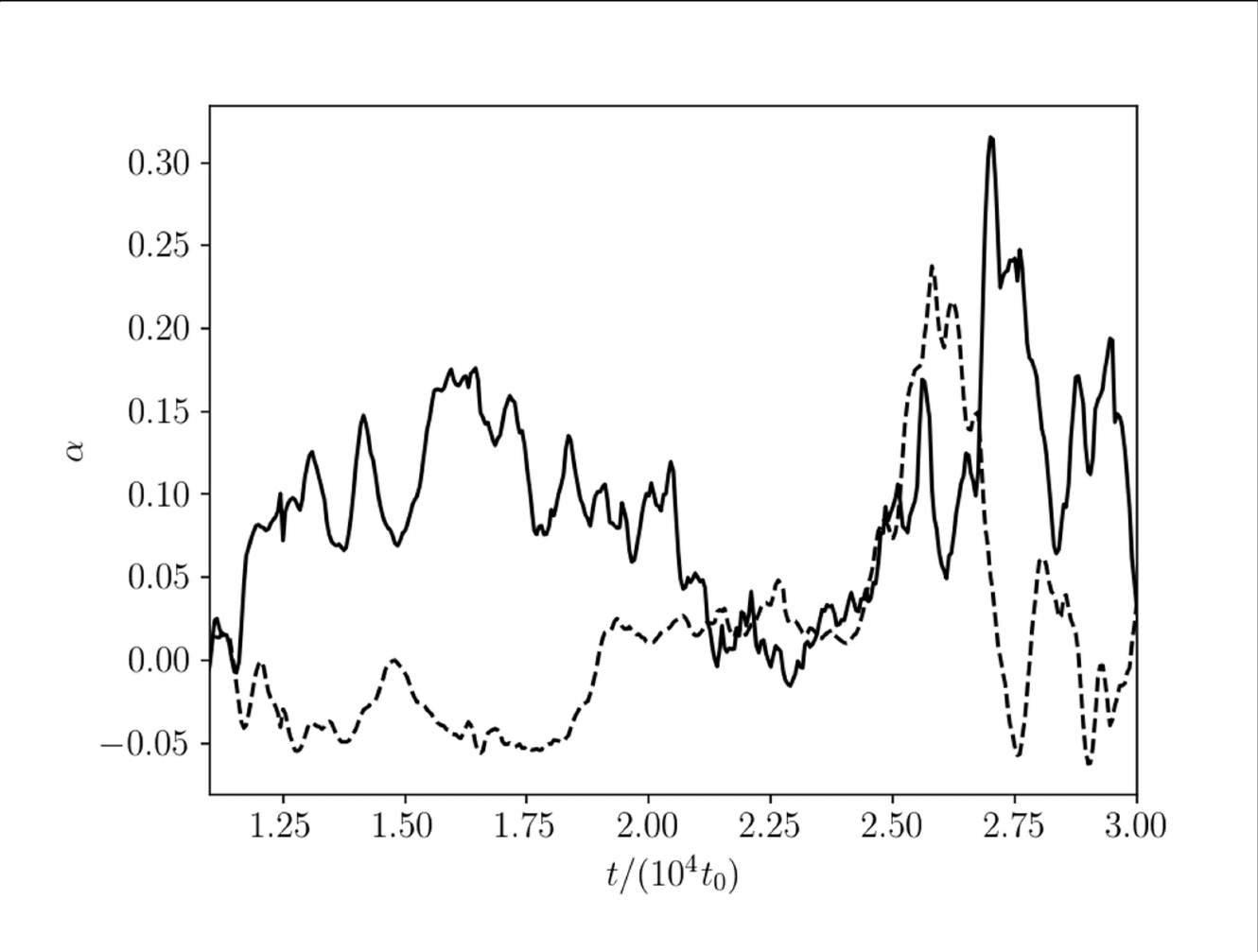}
\caption{{The time evolution of the angular momentum transport rate, $\alpha$ for model M01.
		The solid curve shows $\alpha$ in the surface layer ($0.1<\tau_{\rm es}<1$), and the dashed curve shows $\alpha$ in the disk region ($1<\tau_{\rm es}$).}}
\label{fig11}
\end{figure}
}

\begin{figure}[htbp]
\centering
\plotone{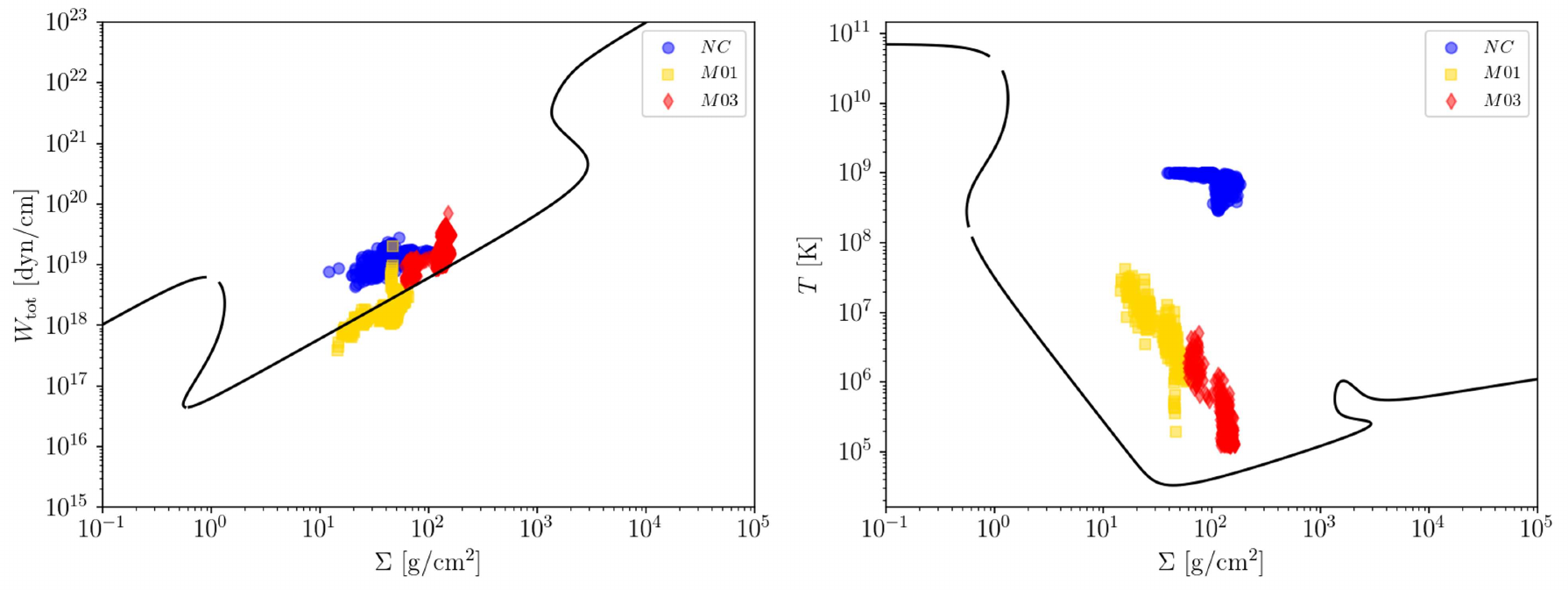}
\caption{The scatter plot of the azimuthally averaged surface density and azimuthally averaged vertically integrated total pressure (left) and density-weighted azimuthally averaged temperature at $z=0$ (right).
		Yellow rectangles and red diamonds show model M01 and M03 at $r=40r_{\rm s}$, respectively.
		Blue circles show the result for model NC at $r=30r_{\rm s}$.
		The parameters for the thermal equilibrium curves are $\Phi_0=3\times10^{16}$ Mx/cm, $\zeta=0.5$, $\Sigma_0=20\ \rm{g/cm^2}$, $\alpha=0.1$, $\alpha'=0$ and $r=40r_{\rm s}$, respectively.}
\label{fig14}
\end{figure}

Figure~\ref{fig14} over plots numerical results and the thermal equilibrium curve when $\alpha=0.1,\ \alpha'=0,\ \Sigma_0=20,\ \Phi_0=2\times10^{16}$, and $\zeta=0.5$.
Here, $\Phi_0$ is the azimuthal magnetic flux, and $\zeta$ is a parameter that relates the magnetic flux and surface density as $\Phi=\Phi_0(\Sigma/\Sigma_0)^\zeta$ (see Appendix).
The yellow rectangles and red diamonds show the surface density and the vertically integrated total pressure for model M01 and model M03 at $r=40r_{\rm s}$.
The blue circles are for model NC at $r=30r_{\rm s}$

Our simulation results are located between the optically thin branch and the optically thick branch.
This is similar to the results for stellar-mass black holes \citep{machida+2006,dexter+2021}.
In model NC (blue circles), $W_{\rm tot}$ is larger than the equilibrium solution because Compton cooling is neglected in this model, and the bremsstrahlung cooling time scale is much longer than the dynamical time scale.

In model M01 (yellow rectangles), the total pressure decreases towards the magnetic pressure-supported disk solution.
The disk contraction enhances the magnetic pressure in the cool region and maintains the disk with a luminosity of around $0.01L_{\rm Edd}$.

In model M03 (red diamonds in Figure~\ref{fig14}), the large optical depth for Thomson scattering enhances the radiation pressure so that the vertically integrated total pressure exceeds that of the magnetic pressure-dominant disk.
The right panel of Figure~\ref{fig14} shows the thermal equilibrium curves in the surface density and temperature planes.
The yellow rectangles and red diamonds show the azimuthally averaged surface density and temperature at $z=0$ for model M01 and model M03 at $r=40r_{\rm s}$ and the blue circles are for model NC at $r=30r_{\rm s}$, respectively.
Our numerical results indicate that the temperature is $10^6-10^7$ K at $\Sigma=20-100\ \rm{g/cm^2}$.
This is consistent with the observation of the soft X-ray emission in CLAGN.
However, the temperature is higher than that in the equilibrium solutions.
The higher temperature can be explained by the additional heating due to the dissipation of magnetic energy {caused by} the merging of the antiparallel azimuthal magnetic field near the equatorial plane.
%

%
{
The additional heating through the injection of the Poynting flux from the inner region can be evaluated by $Q'=(3/2)\alpha W_{\rm tot}(r_{\rm in})\Omega(r_{\rm in})$ where $r_{\rm in}$ is the radius where the outward moving helical flux tubes are formed.
Since $\dot{M}\Omega(r_{\rm in})=2\pi\alpha W_{\rm tot}(r_{\rm in})$, $Q'$ is proportional to $\Omega(r_{\rm in})^2\propto r_{\rm in}^{-3}$ when $\dot{M}$ is fixed.
Thus, at $r=40r_{\rm s}$, $Q'=(3/2)(r/r_{\rm in})^3\alpha W_{\rm tot}(r)\Omega(r)$.
When we denote $Q'=(3/2)\alpha' W_{\rm tot}(r)\Omega(r)$, $\alpha'=(r/r_{\rm in})^3\alpha$.
The heating rate is $Q^+=-\alpha W_{\rm tot} rd\Omega/dr + (3/2)\alpha'W_{\rm tot}\Omega=(3/2)(\alpha+\alpha')W_{\rm tot}\Omega$.
When we adopt $r_{\rm in}=22r_{\rm s}$ as suggested by figure~\ref{fig9}, $\alpha'=(40/22)^3\alpha\sim6\alpha=0.6$.
}

\begin{figure}[htbp]
\centering
\plotone{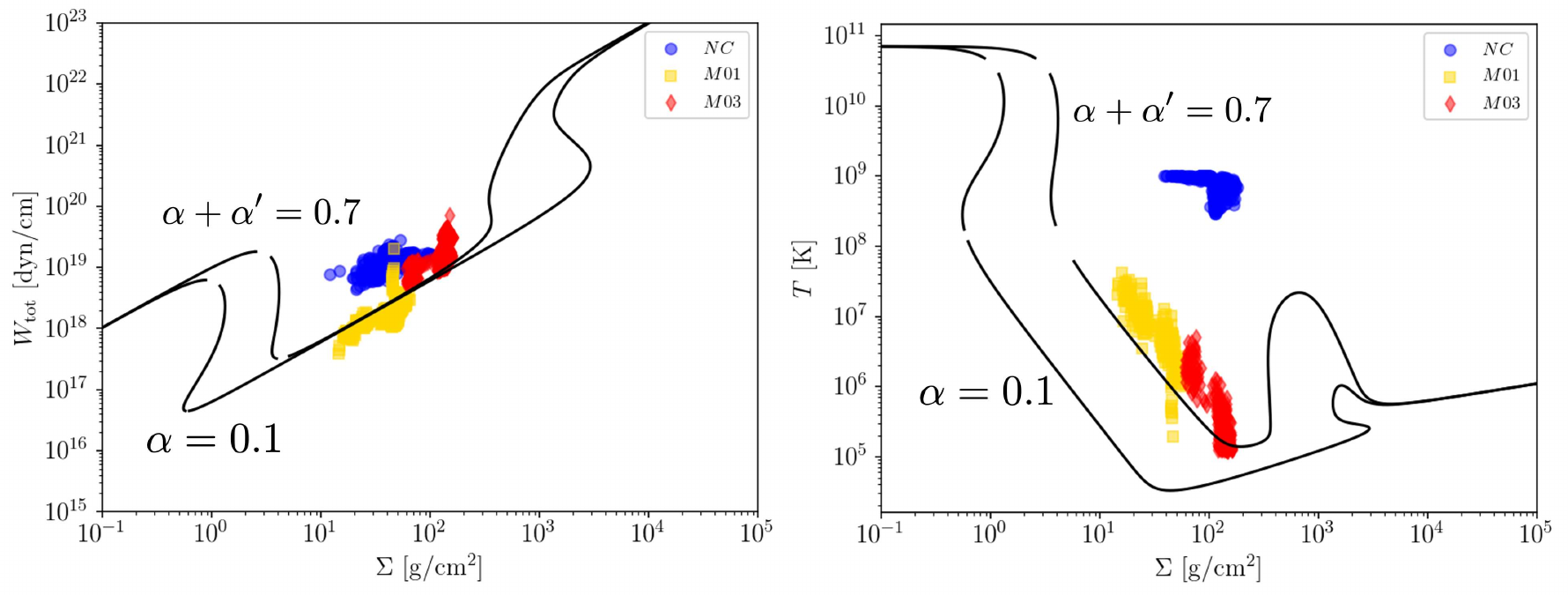}
\caption{Same as figure~\ref{fig15} but for $\alpha=0.1$ and $\alpha+\alpha'=0.7$.}
\label{fig15}
\end{figure}

Figure~\ref{fig15} shows the thermal equilibrium curves in surface density and vertically integrated total pressure (left panel) and temperature (right panel) for $\alpha=0.1$ and $\alpha+\alpha' = 0.7$. 
The yellow rectangles, red diamonds, and blue circles show the numerical results. 
In the left panel of Figure~\ref{fig15}, the magnetic pressure-supported equilibrium solution does not change with $\alpha'$, because the gas pressure does not contribute to the total pressure in this branch (plasma $\beta\sim0.01$ in this regime). 
On the other hand, since the gas temperature increases with the additional heating $(3/2)\alpha'\Omega W_{\rm tot}$, the thermal equilibrium curve for the surface density and temperature approaches the numerical results when $\alpha+\alpha'=0.7$.%

{
Figure~\ref{fig13} schematically shows the energy transport in the warm region suggested by numerical results.
Magnetic field amplification through the surface accretion converts the gravitational energy to the magnetic energy around the interface between the radiatively cooled disk and the inner hot RIAF and forms helical azimuthal magnetic flux tubes.
The helical flux tubes are expelled from the region through Lorentz force toward the positive radial direction and transport the magnetic energy.
The magnetic energy transported to the warm disk is converted to thermal energy through magnetic reconnection.
\begin{figure}[htbp]
\centering
\plotone{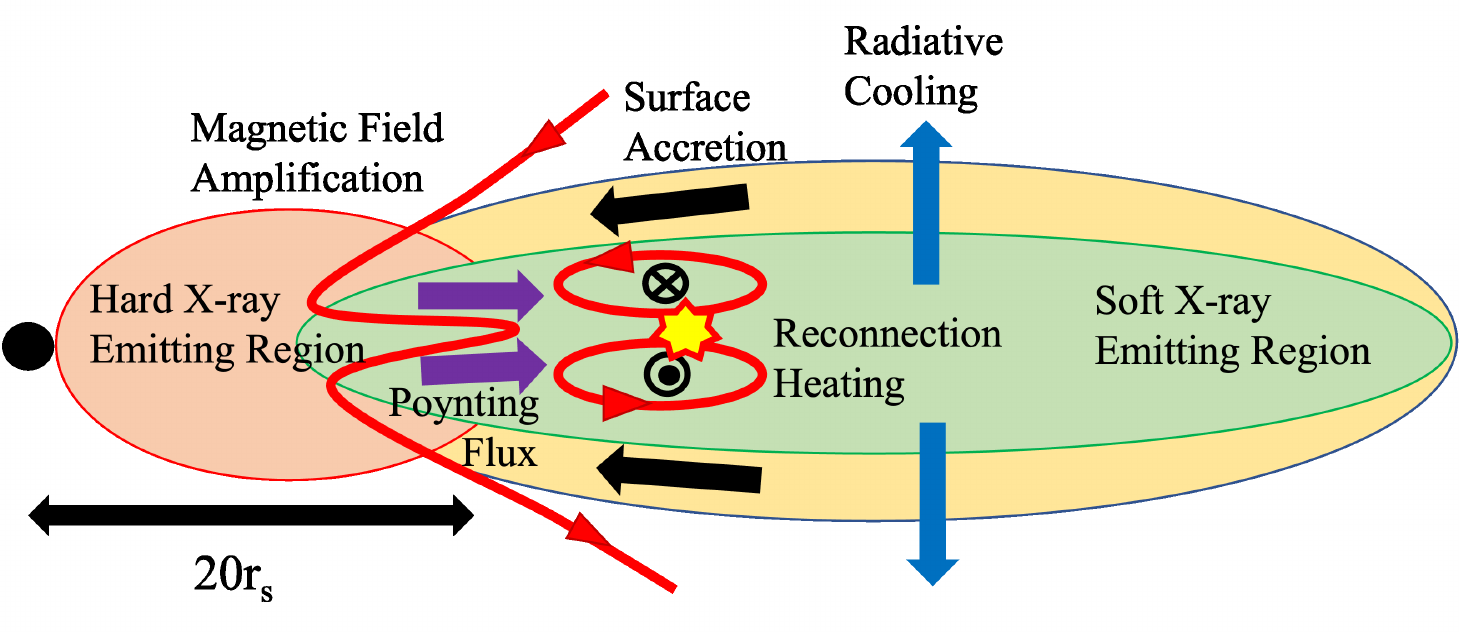}
\caption{{A schematic picture of the mechanism of the magnetic energy transport in the soft X-ray emitting region.
		Red curves show magnetic field lines.
		The surface accretion enhances the magnetic energy.
		The enhanced magnetic energy is transported to the soft X-ray emitting region and converted to thermal energy via the magnetic reconnection.}}
\label{fig13}
\end{figure}
}

It should be noted that the magnetic energy has been accumulated in the disk by mass accretion which releases the gravitational energy.
Therefore, the time-averaged heating rate should be determined by the time-averaged accretion rate.
However, there can be a time lag between the magnetic energy accumulation and the dissipation.
Therefore, there can be a transient state in which the energy dissipation rate is larger than that expected from the accretion rate at that state.
When the magnetic energy release ceases, the cooling will dominate heating, so that the disk will shrink in the vertical direction, which enhances the vertically integrated magnetic energy again.
\section{Summary}
In this paper, we have shown the results of the global three-dimensional radiation MHD simulations with the sub-Eddington accretion rate.
The simulations successfully reproduced the soft X-ray emitting Thomson-thick warm Comptonization region with $T=10^6-10^7\ \rm{K}$.

When the accretion rate is $0.1\dot{M}_{\rm Edd}$ (model M01), a warm, Thomson-thick region with an average temperature of $10^6-10^7$ K is formed outside $40r_{\rm s}$.
In this state, the magnetic pressure is dominant in the warm region.
The bolometric luminosity $L\sim0.01L_{\rm Edd}$ is close to the dark phase of the CLAGN.
As the mass accretion rate increases (model M03), the warm region is formed around $30r_{\rm s}<r<40r_{\rm s}$, and the temperature of the cool outer region decreases to $10^5$ K.
In this state, the cool region ($r>40r_{\rm s}$) is mainly dominated by radiation pressure, because the optical depth is larger than that in model M01.
The luminosity is $\sim0.08L_{\rm Edd}$, which is close to the luminous phase of the CLAGN.
The temperature and the Thomson optical depth of the warm region are consistent with the observational property of the warm Comptonization region in CLAGN  \citep{noda+done2018,tripathi+dewangan2022}.
As the radiation pressure increases further, the radiative cooling rate increases, and the UV-emitting cold region forms which is also consistent with increased UV emission in the luminous state of CLAGN \citep{noda+done2018,tripathi+dewangan2022}.
We should note that the warm region is optically thin for the effective optical depth $\tau_{\rm eff}=\sqrt{\tau_{\rm ff}(\tau_{\rm ff} + \tau_{\rm es})}$ for model M01.
Therefore, soft X-ray emission from reflections of cold disks is ruled out, at least in the low luminosity state.
Soft X-rays are emitted from the warm Comptonization region itself by Comptonizing photons emitted by bremsstrahlung or synchrotron radiation in the same region, or by upscattering photons emitted from the outer cool region by inverse Compton scattering.
In contrast, in model M03, the effective optical depth is close to unity in the $10^5\ \rm{K}$ region.
In this region, soft X-rays can be emitted from the surface region \citep{kawanaka+mineshige2023} as well as the warm Comptonization region near the equatorial plane around $r=30r_{\rm s}$.
We have also obtained the thermal equilibrium solutions of magnetized disks with an azimuthal magnetic field for $10^7M_{\bigodot}$ black holes.
When the magnetic flux at $r=40r_{\rm s}$ exceeds $2\times10^{16}$ Mx/cm, the equilibrium solutions with temperature $T=10^6\sim10^7$ K appear when $\Sigma=10\sim100\ \rm{g/cm^2}$. 
Numerical results show that the accretion flow approaches this state in the $\Sigma-W_{\rm tot}$ plane, but the temperature is an order of magnitude larger than the equilibrium solution unless additional heating is considered.
{
The origin of the additional heating is the magnetic energy dissipated through the reconnection of reversed magnetic fields transported from the inner region. 
}

\begin{acknowledgments}
This work is supported by JSPS Kakenhi 20H01941, 24K00672(PI R.M.), JP23K03448 (PI T.K.), 20K11851, 20H01941, and 20H00156 (H.R.T).
We are grateful to the NAOJ for allowing us to use the XC50 systems operated by CfCA to run the simulations in this study.
We thank Dr. H. Noda, A. Kubota, Y. Kato, S.Yamada, and M. Machida for the valuable discussion.
\end{acknowledgments}

\begin{appendix}

\begin{figure}[htbp]
 	\centering
	\plotone{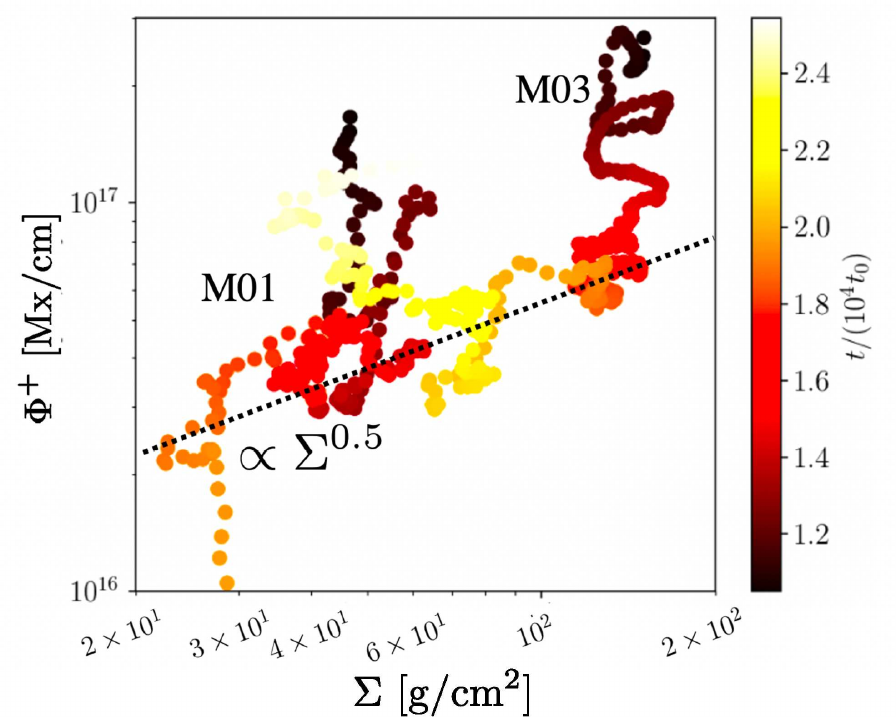}
	\caption{The scatter plot of the azimuthally averaged surface density and the magnetic flux $\Phi_+$ at $r=40r_{\rm s}$ of M01 and M03.
 The dotted line shows $\Phi_+\propto\Sigma^{0.5}$.
 The color shows the time.
		       }
	\label{fig16}
\end{figure}

This Appendix describes the thermal equilibrium solutions for axisymmetric, steady magnetized disks.
Here, we assume the polytropic relation $p_{\rm tot}=K\rho^{(1+1/N)}$, where $N=3$.
Vertical hydrostatic balance yields
\begin{equation}
    \rho(r,z) = \rho_0(r)\left( 1-\frac{z^2}{H^2}\right)^N,
\end{equation}
and
\begin{equation}
    p_{\rm tot}(r,z) = p_{\rm tot,0}(r)\left( 1-\frac{z^2}{H^2} \right)^{N+1},
\end{equation}
where the subscript $0$ denotes the value at the equatorial plane, and $\displaystyle H=\left[ \frac{2(N+1)}{\Omega^2_{K0}}\frac{p_{\rm tot,0}}{\rho_0} \right]^{1/2}$ are the disk half thickness, where $p_{\rm tot}=p_{\rm gas}+p_{\rm rad}+p_{\rm mag}$, and $\Omega_{\rm K}$ are the total pressure, and Keplerian angular velocity, respectively.
Integrating the equations in the vertical direction, the surface density $\Sigma$ and the vertically integrated total pressure $W_{\rm tot}$ can be written as 
\begin{equation}
    \Sigma = \int^{+H}_{-H}\rho dz = 2\rho_0 I_{\rm N}H,
\end{equation}
and 
\begin{equation}
    W_{\rm tot} = \int^{+H}_{-H}p_{\rm tot}dz = 2p_{\rm tot,0}I_{N+1}H,
\end{equation}
where $I_N = (2^NN!)^2/(2N+1)!$.
{By rewriting $\rho_0$ and $p_{\rm tot}$ using $\Sigma$ and $W_{\rm tot}$, we obtain}
\begin{equation}
    H = \left( \frac{2N+3}{\Omega^2_{\rm K0}}\frac{W_{\rm tot}}{\Sigma} \right)^{1/2}.
\end{equation}
Now, assuming axisymmetry and  integrating the azimuthally averaged equations in the vertical direction, we get:
\begin{equation}
    \dot{M} = -2\pi r\Sigma v_r = \rm{const.},
\end{equation}
\begin{equation}
    \dot{M}(l-l_{\rm in}) = 2\pi r^2\alpha W_{\rm tot},
\end{equation}
\begin{equation}
    \frac{\dot{M}}{2\pi r^2}\frac{W_{\rm rad}+W_{\rm gas}}{\Sigma}\xi = Q^+ - Q^-,
    \label{eq:ene}
\end{equation}
where $\ell$ is the specific angular momentum and $\ell_{\rm in}$ is the specific angular momentum carried into the black hole.
The left-hand side of equation~(\ref{eq:ene}) is the advective cooling rate $Q_{\rm adv}$.
We assume $\xi=1$.
\citet{oda+2009} assumed that the magnetic flux advection rate $\Phi v_r$ {where $\Phi=\int B_\varphi dz$} is determined by $\dot{M}$.
Here we modify the formulation by introducing a parameter $\zeta$ and assume that 
\begin{equation}
    \Phi = \Phi_0\left( \frac{\Sigma}{\Sigma_0} \right)^\zeta.
\end{equation}
When $\zeta>0$, the magnetic flux is stored in the region where the mass accumulates.
Here we should note that the magnetic flux $\Phi$ is the magnetic flux integrated in the vertical direction.
Thus the azimuthal magnetic field cancels when the azimuthal magnetic field is anti-symmetric with respect to the equatorial plane. 
Here we use $|B_\varphi|$ to avoid cancellation of the azimuthal magnetic flux when $B_\varphi$ is anti-symmetric with respect to the equatorial plane.
Figure~\ref{fig16} shows the relation between the absolute magnetic flux $\Phi_+=\int|B_\varphi|dz$ and $\Sigma$ at $r=40r_{\rm s}$ obtained by radiation MHD simulations.
The magnetic flux is larger in the early stage.
In the later stage, the magnetic flux decreases, because the anti-symmetric azimuthal magnetic field dissipates around the equatorial plane.
The azimuthal magnetic flux $\Phi_0$ and the parameter $\zeta$ can be estimated to be $\Phi_0=2\times10^{16}\ \rm{Mx/cm}$ and $0.5$, respectively.

We further assume that the disk heating rate is written as 
\begin{equation}
    Q^+ = -\alpha W_{\rm tot}r\frac{d\Omega}{dr} + \frac{3}{2}\alpha'W_{\rm tot}\Omega.
\end{equation}
Here, the first term on the right-hand side is the heating by the vertically integrated $r\varphi$ component of the stress tensor $W_{\rm r\varphi}=-\alpha W_{\rm tot}$ and the second term is the enhanced heating by magnetic reconnection, where $\alpha'$ is proportional to the reconnection rate. 

For radiative cooling, we consider thermal bremsstrahlung emission.
The cooling rate in the optically thin limit is written as 
\begin{equation}
	Q^-_{\rm thin} = 6.2\times10^{20}\frac{I_{2N+1/2}}{2I_N^2}\frac{\Sigma^2}{H}T_0^{1/2},
\end{equation}
and  the cooling rate in the optically thick limit is written as 
\begin{equation}
	Q^-_{\rm thick} = \frac{16\sigma I_NT_0^4}{3\tau/2},
\end{equation}
where $\sigma$ and $\tau=\tau_{\rm abs}+\tau_{\rm es}$ are Stefan-Boltzmann constant and the total optical depth.
Here $\tau_{\rm abs}$ is the absorption optical depth, and $\tau_{\rm es}=0.5\kappa_{\rm es}\Sigma$ is the optical depth for Thomson scattering.

For the intermediate case, we use the approximate form of radiative cooling \citep[e.g.,][]{abramowicz+1996},
\begin{equation}
	Q^- = \frac{16\sigma I_N T^4_0}{3\tau/2 + \sqrt{3} + \tau_{\rm abs}^{-1}},
\end{equation}
where 
\begin{equation}
	\tau_{\rm abs} = \frac{8I_{\rm N}^2}{3I_{\rm N+1}\tau}\frac{Q^-_{\rm thin}}{Q^-_{\rm thick}}.
\end{equation}

We also consider the cooling by the inverse Compton scattering.
The cooling rate by the inverse Compton scattering can be written as 
\begin{equation}
	Q^-_{\rm Comp} = \kappa_{\rm es}\Sigma Q^-\frac{4k_{\rm B}}{m_{\rm e}c^2}\left( \frac{I_{\rm N+1}}{I_{\rm N}}T_0 - T_{\rm r}\right),
\end{equation}
where 
\begin{equation}
	T_{\rm r}=\left(\frac{3\tau/2}{4a_{\rm r}cI_{\rm N}}Q^-\right)^{1/4}
\end{equation}
is the radiation temperature.
For simplicity, we assume that the ion and electron temperatures are the same.
However, we should consider the temperature difference between ions and electrons in hot accretion flows where $T_i>10^9$ K.
We can neglect the temperature difference between electrons and ions in the warm Comptonization region where $T_{i}<10^8$ K. 

To obtain the thermal equilibrium curves of the black hole accretion flows, taking into account the azimuthal magnetic field we need to solve the equation of state and the balance between heating, radiative cooling, and advective cooling.
Defining $f_1 = W_{\rm tot} - W_{\rm rad} -W_{\rm gas} - W_{\rm mag}$ and $f_2 = Q^+ - Q^-_{\rm rad} - Q^-_{\rm adv}$, we obtain
\begin{equation}
	f_1 = W_{\rm tot} - \frac{1}{4c}  \frac{16\sigma I_{\rm 3}T_0^4}{3\tau/2 + \sqrt{3} + \tau_{\rm abs}^{-1}} \frac{I_{\rm N+1}}{I_{\rm N}}H\left(\tau + \frac{2}{\sqrt{3}}\right) - \frac{I_{\rm N+1}}{I_{\rm N}}\frac{R}{\mu}\Sigma T_0 - \frac{\Phi_0^2}{8\pi H}\left(\frac{\Sigma}{\Sigma_0}\right)^{2\zeta},
\end{equation}
and 
\begin{equation}
	f_2 = Q^+ - \frac{16\sigma I_{\rm N}T_0^4}{3\tau/2 + \sqrt{3} + \tau_{\rm abs}^{-1}} - \kappa_{\rm es}\Sigma Q^-\frac{4k_{\rm b}}{m_{\rm e}c^2}\left( \frac{I_{\rm N+1}}{I_{\rm N}}T_0 - T_{\rm r}\right) -\frac{\dot{M}}{r^2}\frac{W_{\rm gas} + W_{\rm rad}}{\kappa_{\rm es}\Sigma}\xi.
\end{equation}
The thermal equilibrium curve is obtained by solving $f_1 = f_2 = 0$ for a given radius and mass accretion rate.

\begin{figure}[htbp]
 	\centering
	\plotone{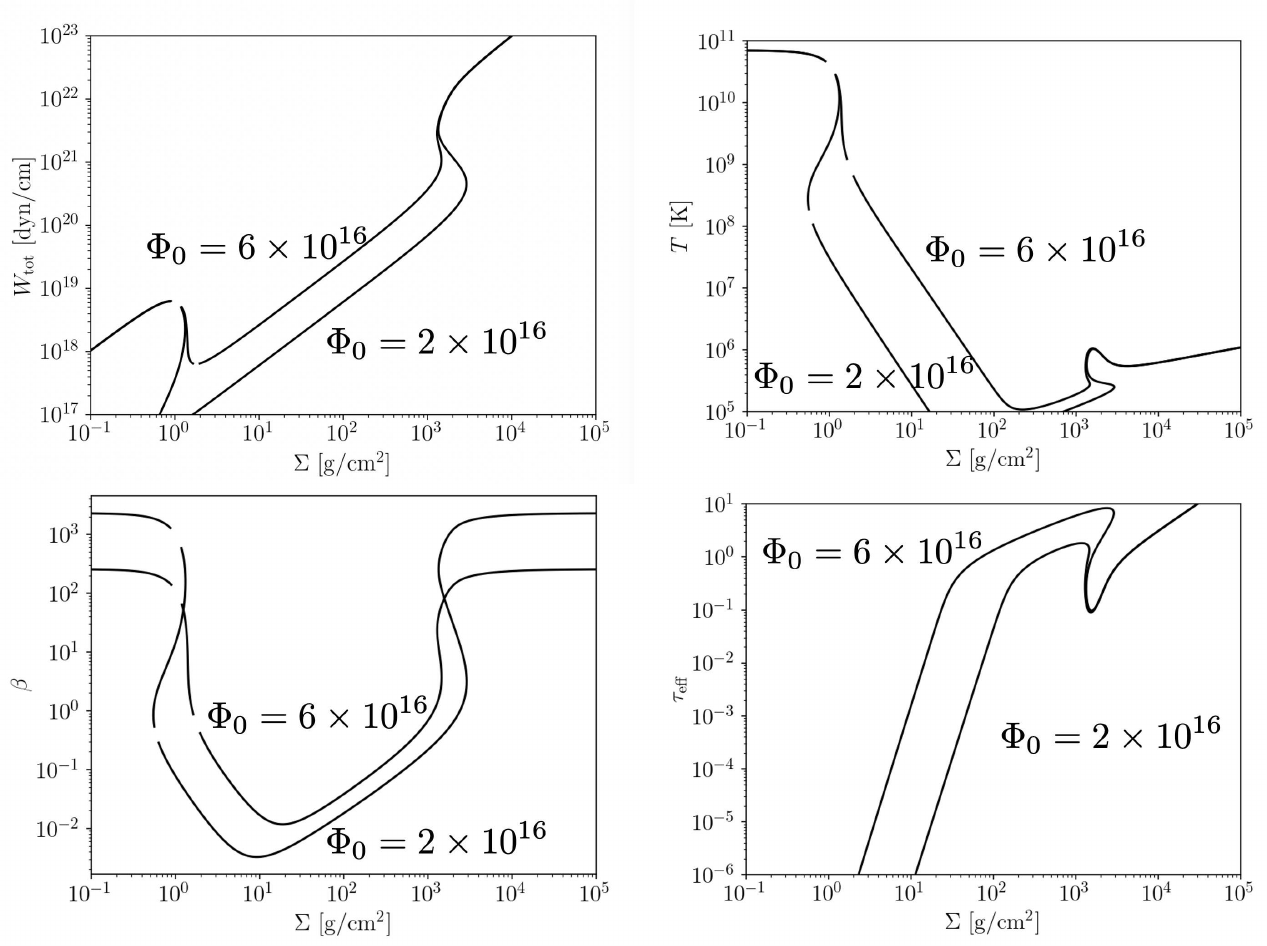}
	\caption{Thermal equilibrium curves for disks partially supported by azimuthal magnetic fields at $r=40r_{\rm s}$, $\alpha=0.1$ and $\alpha'=0$ in the plane of surface density and $W_{\rm tot}$ (upper left), temperature (upper right), plasma $\beta$ (lower left), and effective optical depth $\tau_{\rm eff}=\sqrt{\tau_{\rm abs}\tau_{\rm es}}$ (lower right).
		       The black hole mass, $\Sigma_0$, and $\zeta$ are assumed to be $10^7M_{\bigodot}$, $30\ \rm{g/cm^2}$, and $0.5$, respectively.
		       The magnetic flux for each curve is noted in the figure.
		       }
	\label{fig17}
\end{figure}

Figure~\ref{fig17} shows the result for a supermassive black hole with mass $M=10^7M_{\bigodot}$ at $r=40r_{\rm s}$ when $\zeta=0.5$, $\alpha=0.1, \alpha'=0$, and $\Sigma_0=30\ \rm{g/cm^2}$.
The upper left panel of Figure~\ref{fig17} shows the solution for the vertically integrated total pressure and surface density.
When the surface density exceeds the upper limit for RIAF, the vertically integrated total pressure decreases because the radiative cooling overcomes the disk heating.
However, since the magnetic pressure increases as the disk shrinks in the vertical direction, the enhanced heating balances the radiative cooling, so that the intermediate state between the optically thin branch and the optically thick standard disk appears as shown in \citet{oda+2009} for a stellar-mass black hole.

The upper right panel of figure~\ref{fig17} shows the thermal equilibrium curve in the surface density and $W_{\rm tot}$ planes.
When the magnetic flux exceeds $2\times10^{16}\ $Mx/cm, a warm ($T=10^6-10^7$ K) region appears when $\Sigma=1\sim100\ \rm{g/cm^2}$.
The temperature and Thomson optical depth in this region are comparable to the warm Comptonization region and can explain the soft X-ray emission.
Note that the temperature of this region is lower than that of stellar-mass black holes, where $T=10^8$ K \citep{oda+2009,oda+2012}.

The lower left panel of figure~\ref{fig17} shows the plasma $\beta$ defined as $\beta=(W_{\rm gas}+W_{\rm rad})/W_{\rm mag}$.
In the intermediate state, the magnetic pressure dominates the gas and radiation pressure.
Note that the plasma $\beta$ is lower for the lower magnetic flux model.
This is because the disk heating rate can only balance the radiative cooling when the magnetic pressure is much higher than the gas pressure.

The bottom right panel shows that the warm Comptonization region is optically thin for the effective optical depth.

\end{appendix}

\bibliography{ms}
\bibliographystyle{aasjournal}



\end{document}